# The impacts of artificial intelligence on environmental sustainability and human well-being

Noemi Luna Carmeno[1*], Tiago Domingos[2], Daniel W. O'Neill[1,3*]

[1]UB School of Economics, Universitat de Barcelona, Barcelona, Spain
[2]MARETEC – Marine, Environment and Technology Center, LARSyS – Laboratory for Robotics and Engineering Systems, Instituto Superior Técnico, Universidade de Lisboa, Lisbon, Portugal
[3]Sustainability Research Institute, School of Earth and Environment, University of Leeds, Leeds, UK
[*]Corresponding authors: noemi.luna.carmeno@ub.edu; oneill@ub.edu

## Abstract

Artificial intelligence (AI) is increasingly being described as a transformative general-purpose technology, yet its impacts on environmental sustainability and human well-being remain poorly understood. Here, we conduct a systematic review of 1,291 studies selected from 6,655 records to map how the literature assesses these impacts. We find that current research provides a fragmented and uneven account of AI's consequences for human well-being and the environment. Environmental studies focus narrowly on energy use and $CO_2$ emissions (72%) and rarely consider systemic effects (11%), while well-being studies are predominantly conceptual and overlook subjective well-being, cognitive capabilities, and upstream supply-chain impacts. Strikingly, 82% of environmental studies portray AI's impacts as positive, but this finding is driven by the large number of application-level studies. Well-being analyses show a near-even split (44% positive; 46% negative). However, this split masks differences across well-being dimensions: impacts on income and health are generally expected to be positive, whereas impacts on inequality, social cohesion, and employment are expected to be negative. Based on our findings, we identify important priorities for future research: environmental assessments should consider systemic effects and indicators beyond energy and $CO_2$, while well-being research should prioritise empirical analysis. More fundamentally, research must consider environmental sustainability and human well-being together, recognising that AI's impacts are deeply interconnected. Towards this aim, we extend an existing three-level framework for the climate impacts of AI to include broader environmental impacts and human well-being, providing a unified basis to assess AI impacts and guide its development towards environmental sustainability and human flourishing.

## 1. Introduction

With the widespread adoption of large language models (LLMs) since 2022, AI is increasingly being viewed as a general-purpose technology, with transformative effects comparable to electricity or the steam engine (Eloundou et al. 2024; Goldfarb et al. 2023). Yet, the scale and nature of these transformations remain poorly understood. As AI becomes more embedded in economic and social systems, concerns are growing about its potential environmental footprint and societal consequences. These concerns are particularly relevant in light of escalating sustainability challenges, such as the transgression of planetary boundaries (Sakschewski et al. 2025), and social problems, including rising inequality, political polarisation, and declining youth well-being (Blanchflower 2025; Chancel et al. 2025; Nord et al. 2025). It remains unclear how AI will affect many of these trends.

AI impacts are not only determined by the technical capabilities of models. They also arise from the interactions among models, data, infrastructure, organisations, institutions, and the social, economic, and political contexts in which AI is developed and deployed (Kudina and van de Poel 2024; Peng et al. 2021). As a consequence, impacts occur simultaneously at multiple levels (Kaack et al. 2022). Benefits at the application level may coexist with upstream environmental and social burdens or with broader systemic effects, while both benefits and costs may be distributed unevenly across populations and regions (Kaack et al. 2022; Lehuedé 2025; Varoquaux et al. 2025). Environmental sustainability and human well-being impacts are also closely interconnected, as environmental pressures affect people and communities, while social, economic, and political changes influence resource use and environmental outcomes (Dong et al. 2024; Gaffney et al. 2025; Luccioni et al. 2025a).

Research on AI's environmental and well-being impacts has grown substantially in recent years but remains piecemeal and siloed, with limited synthesis of findings and gaps. Existing reviews typically focus on specific impacts, including energy use, employment, and health (Kamiya and Coroamă 2025; Martinez-Millana et al. 2022; Peppiatt 2024), particular industries or applications, such as construction and sustainable plastics (Guarda et al. 2024; Regona et al. 2024), or individual disciplinary perspectives, including ethics and economics (Chuang et al. 2022; Hagendorff 2024; Lu and Zhou 2021). Moreover, reviews generally consider either environmental or well-being dimensions in isolation (Malja and Afrasiabi 2026; Pachegowda 2023; Verdecchia et al. 2023).

This fragmentation limits our understanding of which impacts the literature considers, how they are assessed, and where the main gaps lie. It also obscures connections and trade-offs between environmental and well-being impacts, as well as how the immediate effects of specific applications interact with upstream and broader systemic impacts. To our knowledge, no study to date has considered the literature on the impacts of AI on the environment and well-being together. Bringing these research fields together is essential to compare their scope, methods, and conclusions, identify overlooked impacts and interactions, and inform future research and policy through a more integrated evidence base on AI's impacts on people and the planet.

Against this backdrop, we address four questions: (i) Which environmental sustainability and human well-being impacts of AI are considered in the literature? (ii) How are these impacts assessed? (iii) How do the environmental and well-being literatures compare? (iv) What research gaps and priorities for future work emerge?

To answer these questions, we systematically review 1,291 studies published from 2010 onwards, selected from 6,655 records. We adopt a broad, multidimensional understanding of AI's impacts. With respect to the environment, we examine energy use, $CO_2$ emissions, water use, material use, land use, ecological footprint, biodiversity, environmentally-related Sustainable Development Goals (SDGs), and other pollutants. With respect to human well-being, we examine health, employment, income, inequality, social cohesion, subjective well-being, cognitive capabilities, and existential risk.

We combine quantitative mapping with qualitative synthesis, systematically compare the environmental and well-being literatures, and propose a unified framework for analysing both dimensions across computing-related, application, and systemic levels. Based on this analysis, we identify research gaps and outline priorities for future research.

## 2. Methods

We conducted a systematic literature review in accordance with the PRISMA 2020 framework (Page et al. 2021). Separate searches for environmental sustainability and human well-being impacts were developed iteratively and applied across Scopus, arXiv, and National Bureau of Economic Research (NBER) Working Papers. The searches covered publications up to 31 December 2024 and were supplemented by backward and forward citation tracking to capture recent and relevant contributions across environmental sciences, social sciences, computer science, engineering, medicine, philosophy, and economics. Studies were included if they: (i) were published in English from 2010 onwards; (ii) were peer-reviewed articles, working papers, preprints, conference proceedings, or grey literature from reputable institutions; (iii) had full text available; and (iv) addressed one of our four research questions.

After duplicate records were removed, title/abstract and full-text screening were conducted blind to authorship and publication source to reduce selection bias. Included studies were classified into three relevance categories. Studies were classified as very relevant if their core contribution directly addressed one or more research questions, relevant if they addressed a research question but it was not their main contribution, and less relevant if their analysis intersected with a research question only tangentially. Methodological quality was assessed during full-text screening by examining the clarity of assumptions, suitability of the analytical approach, and coherence between the evidence, methods, and conclusions. Studies judged to be at critical risk of bias were excluded. Screening decisions and exclusion reasons were recorded in a structured spreadsheet to provide an auditable record of the selection process.

Data were systematically extracted into a spreadsheet using an iteratively developed coding framework. The impact categories were identified inductively from the literature during data extraction. Related impacts were grouped, definitions were refined, and the taxonomy was expanded as additional dimensions emerged, until a comprehensive and internally consistent classification was achieved. The extraction form combines categorical fields for quantitative mapping with free-text fields covering findings, methods, assumptions, limitations, and research gaps. For very relevant and relevant studies, data were extracted manually. For studies classified as less relevant, comprising approximately 30% of the sample, data were extracted with the support of an LLM (Microsoft 365 Copilot Enterprise) using standardised prompts. Every extracted item and category assignment was manually checked against the original study before entry into the final dataset, and no unverified outputs were retained.

We present both quantitative and qualitative results from our review. The quantitative analysis maps the distribution of studies by type of environmental or well-being impact considered (e.g. energy use, $CO_2$ emissions, health, employment), level at which impacts arise (computing-related, application-level, systemic), broad method used (conceptual/theoretical, qualitative, quantitative), specific method used (e.g. statistical analysis, optimisation), analytical scale (micro or macro), and dominant sentiment (positive, neutral, negative). The three levels at which impacts arise are adapted from Kaack et al.'s (2022) framework for AI-related greenhouse gas emissions, which we extend to include other environmental impacts and subsequently apply to human well-being. Sentiment was assigned based on the balance of positive and negative impacts reported by the authors of each study. If no overall tendency was discernible, the study was classified as neutral.

The qualitative analysis synthesises recurring findings, highlights major research gaps, and proposes a research agenda to guide future work. For well-represented topics we synthesise convergent findings, while for understudied topics we report available findings and key limitations. Research gaps were identified based on explicit mentions in the literature as well as from our quantitative analysis. The quantitative and qualitative findings were then used to develop a unified framework that integrates and organises AI's environmental and well-being impacts according to the level at which they arise (computing-related, application-level, or systemic).

The full review protocol is available on protocols.io (Carmeno et al. 2026). Complete search strings, operational definitions, the PRISMA flow diagram (Supplementary Fig. 1), detailed extraction procedures, and additional reproducibility information are provided in the Supplementary Information to this article.

## 3. Quantitative results

Of the 1,291 studies included in our review, 523 address environmental impacts and 768 address impacts on human well-being. Overall, the literatures on the environmental and well-being impacts of AI differ markedly in terms of their outlook, methods, analytical scale, and the levels at which impacts are considered. Environmental studies predominantly suggest positive effects, centre on energy use and $CO_2$ emissions, and are mostly quantitative and micro in scale, with limited attention to systemic implications. By contrast, the well-being literature is more balanced in its overall sentiment, although sentiment differs markedly by impact type, from income and health (most positive) to existential risk and inequality (most negative). Moreover, the well-being literature is predominantly conceptual, with limited empirical evidence. It tends to be macro in scale and to focus on societal-level impacts such as social cohesion, inequality, and employment. Below we explore our specific findings.

### 3.1 Energy use and $CO_2$ emissions dominate research on the environmental impacts of AI

Environmental studies are very heavily concentrated on energy use and $CO_2$ emissions, which together represent 72% of the sample (Fig. 1a). Broader indicators such as the SDGs or composite indicators such as the ecological footprint are considered much less frequently. Despite being a growing concern (Li et al. 2025), water use is addressed in less than 5% of studies. Material use, biodiversity, and land use are rarely considered. The prominence of energy use and $CO_2$ emissions in the literature is reinforced by the fact that $CO_2$ is the main driving component of the ecological footprint indicator. Studies coded as "other pollutants" mainly address non-$CO_2$ greenhouse gases.

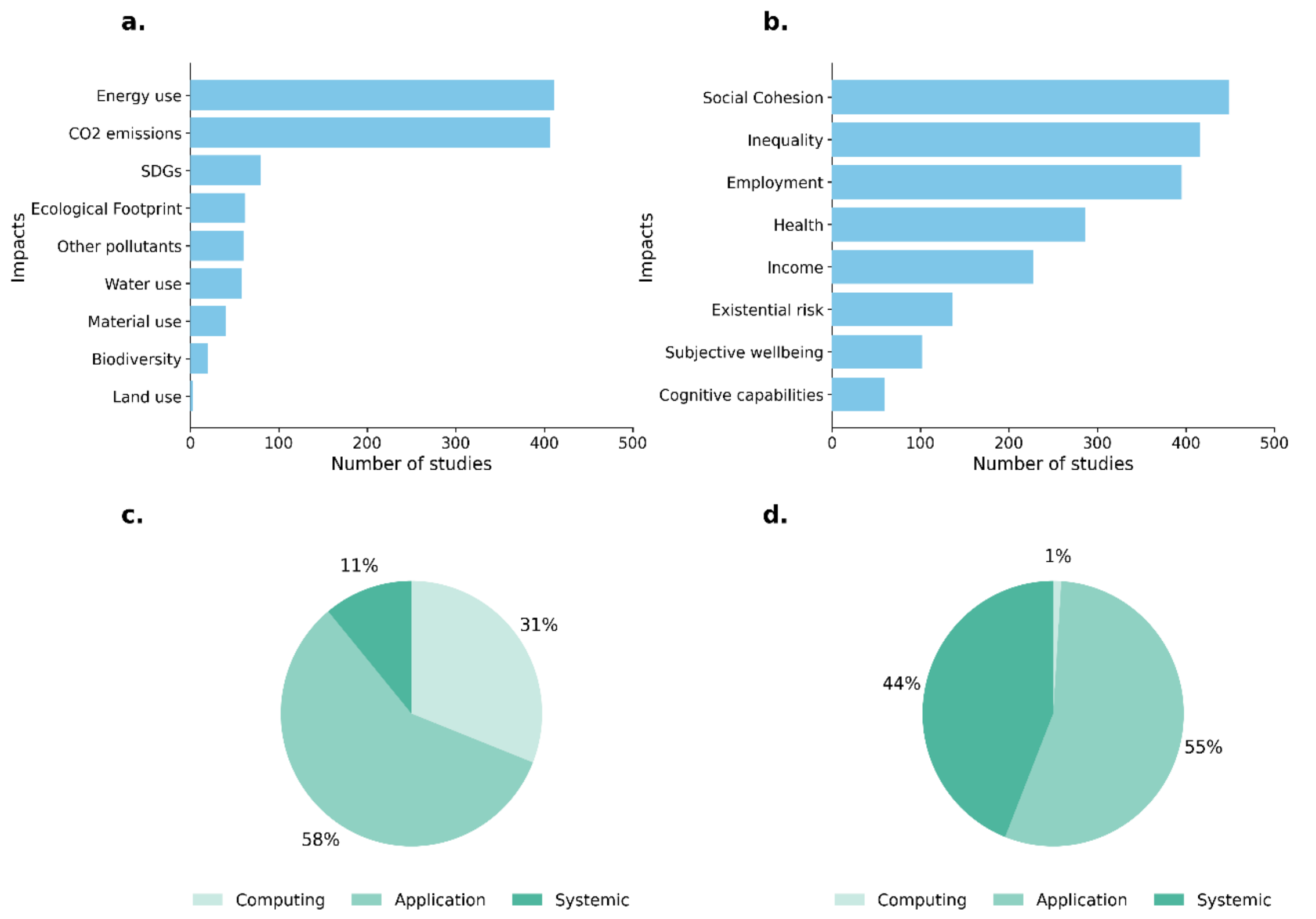

**Fig. 1 Distribution of studies by impact type and level at which impacts arise. (a)** Environmental studies by impact type. **(b)** Well-being studies by impact type. **(c)** Environmental studies by level at which impacts arise (computing-related, application-level, or systemic). **(d)** Well-being studies by level at which impacts arise. Environmental studies are concentrated on energy use, $CO_2$ emissions, and application-level impacts, whereas well-being studies cover a broader range of social impacts but rarely address cognitive capabilities, subjective well-being, or computing-related impacts

### 3.2 Cognitive impacts and subjective well-being are generally overlooked in social impact research on AI

Well-being research is more evenly distributed across impact types. The most considered impacts are social cohesion, inequality, and employment, followed by health and income (Fig. 1b). In contrast, cognitive impacts and subjective well-being are hardly addressed. Research on social cohesion impacts, including concentration of power, misinformation, surveillance, and the weakening of trust and community bonds, has recently become the most prevalent.

### 3.3 Systemic impacts are underexplored with respect to the environment

Building on a helpful framework proposed by Kaack et al. (2022), the environmental impacts of AI can be grouped into three categories based on the level at which they occur: *computing-related impacts* (i.e. resource use and pollution from model training, inference, and the hardware life-cycle); *application-level impacts* (i.e. immediate effects of specific uses); and *systemic impacts* (i.e. indirect, long-term consequences arising from broader changes in behaviour or institutions, such as rebound effects). Our analysis reveals that most environmental studies focus on impacts at the application level (58%), while only 11% address systemic impacts (Fig. 1c). The limited focus on systemic impacts risks overlooking wider consequences that could outweigh or counteract application-level impacts.

### 3.4 Social impacts arising from the AI upstream supply chain are underexamined

Applying the same three-level categorisation to well-being studies shows that most studies address application-level (55%) and systemic effects (44%), while computing-related social impacts receive very limited attention (only 1%; Fig. 1d). Where assessed, computing-related impacts include uncompensated data provision, precarious data labour concentrated in the Global South, community burdens from material extraction and infrastructure siting, and distributional effects that may reinforce unequal exchange. Limited consideration of computing-related social impacts may lead current assessments to underestimate the extent to which AI supply chains reinforce inequalities between the Global North and Global South in the distribution of labour, environmental and social burdens, and economic benefits.

### 3.5 Environmental and well-being studies differ sharply in methods and scope

Research on AI's impacts on the environment and on human well-being diverges markedly in terms of both methods and scale of analysis. Environmental impact studies are predominantly quantitative (62%; Fig. 2a) and largely micro in scale (55%; Fig. 2b). This finding is largely driven by the quantitative nature of computing-related and application-level studies (Fig. 2c).

By contrast, well-being studies are primarily conceptual (54%), with fewer empirical contributions (18% qualitative and 28% quantitative; Fig. 2a). Well-being studies also tend to adopt a more macro perspective (54%; Fig. 2b). The lack of empirical work is particularly pronounced for some impact types, namely existential risk (86% conceptual; Fig. 2d), health (66% conceptual), and social cohesion (64% conceptual).

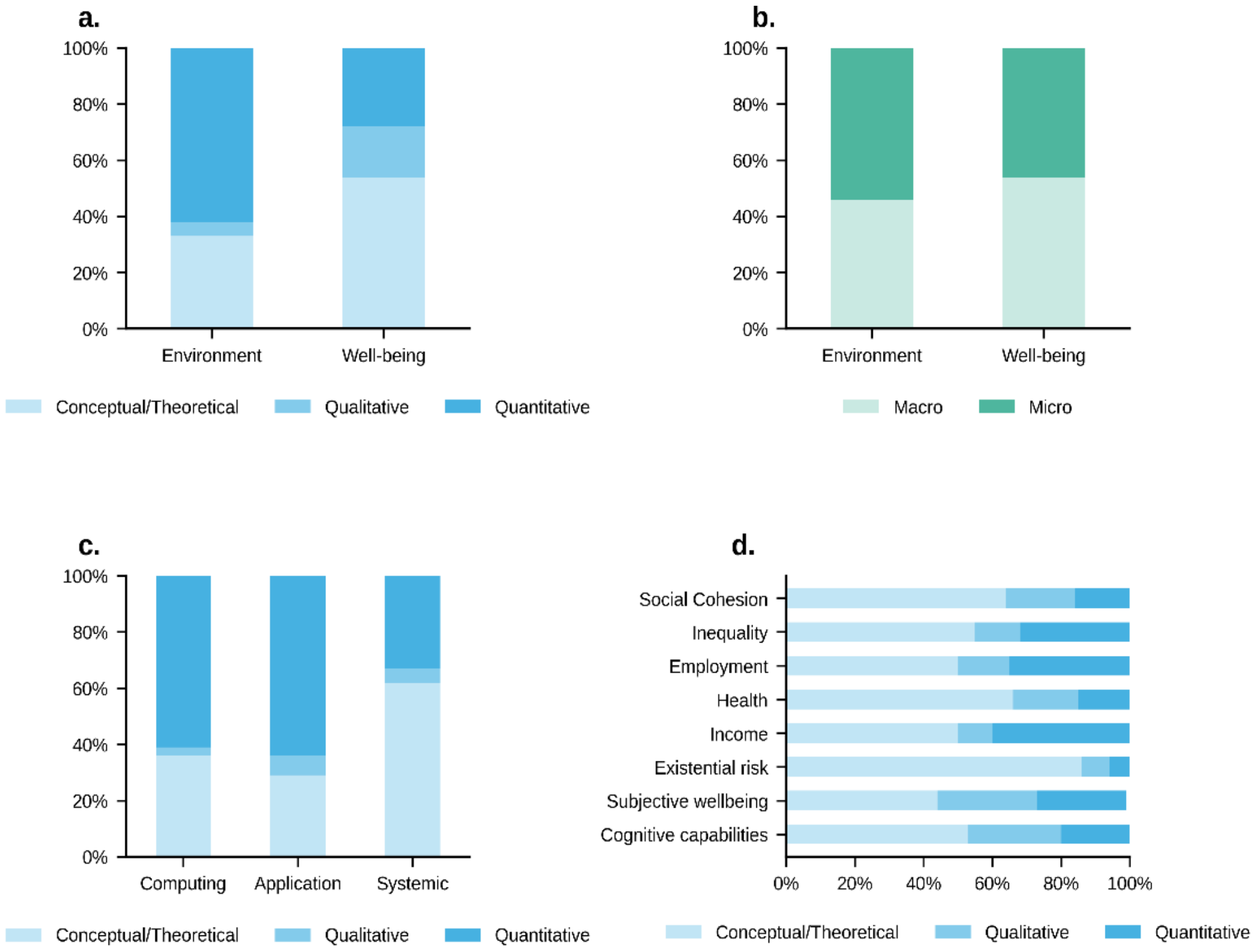

**Fig. 2 Percentage of studies by broad method used and analytical scale. (a)** Percentage of environmental and well-being studies by broad method used, classified as conceptual/theoretical, qualitative, or quantitative. **(b)** Percentage of environmental and well-being studies by scale of analysis, distinguishing micro studies (application or enterprise) from macro studies (national or global). **(c)** Percentage of environmental studies by the level at which impacts arise (computing-related, application-level, or systemic) and broad method used. **(d)** Percentage of well-being studies by impact type and broad method used

### 3.6 Both environmental and well-being studies tend to rely on a narrow set of methods

Examining impact types alongside a more granular classification of methods, environmental impact research appears tightly clustered around a narrow set of approaches and impact types (Fig. 3a). Environmental studies typically employ optimisation models for energy efficiency, econometric analyses, case studies, or experimental measurements (e.g. carbon trackers) focusing on energy use and $CO_2$ emissions. Specialised techniques, such as remote sensing to assess data-centre footprints, remain rare.

Well-being impact research is more evenly distributed in terms of impact types but still tends to concentrate around a few conceptual methods, such as conceptual frameworks, literature reviews, ethical analysis, and policy analysis (Fig. 3b). Statistical and macroeconomic analyses are also relatively well-represented, particularly regarding employment, inequality, and income. By contrast, participatory approaches are underrepresented. Such approaches could help engage affected stakeholders and provide contextual insights into AI's effects on social groups and communities (Fraisl et al. 2025; Wang et al. 2024c).

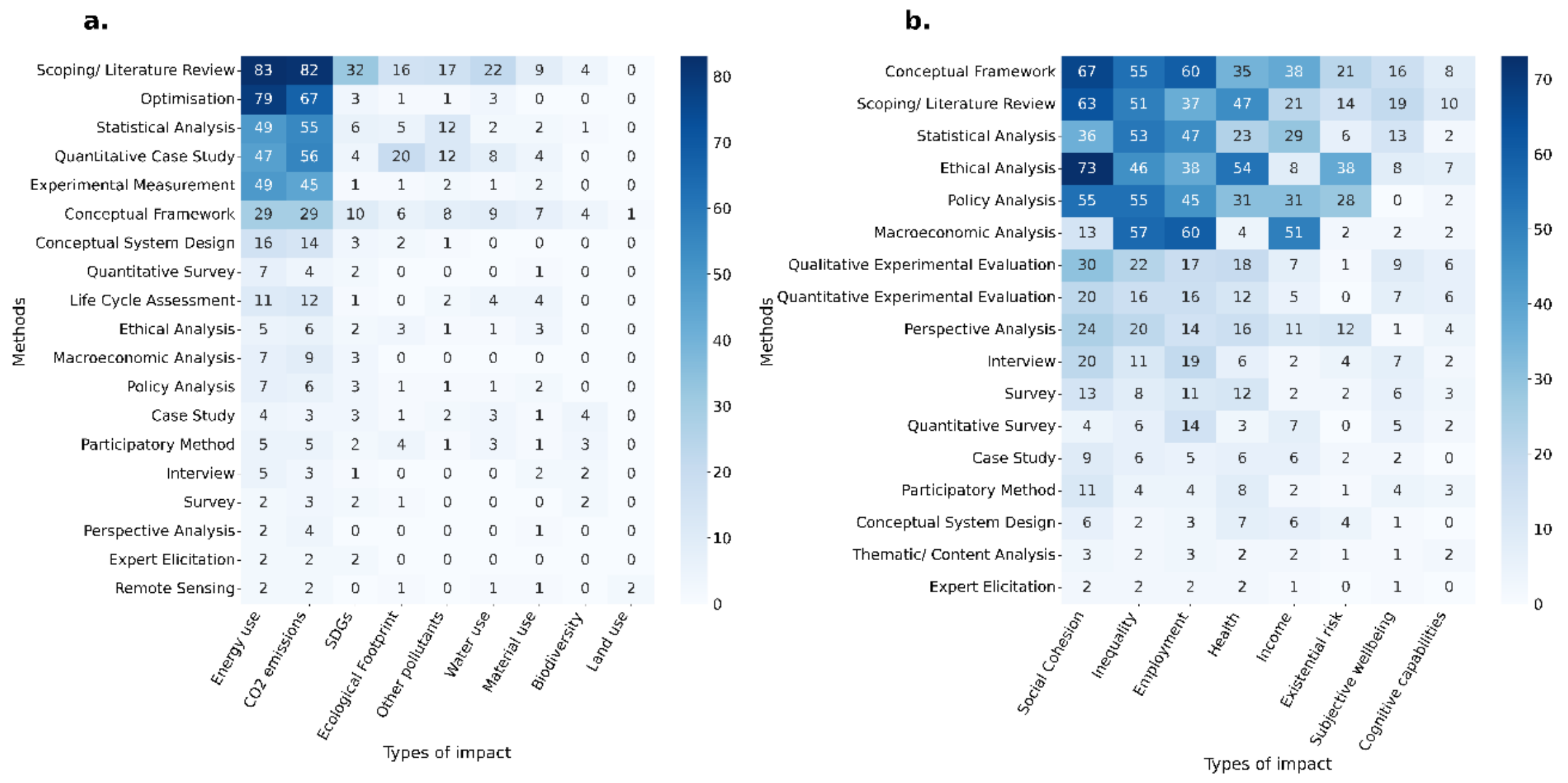

**Fig. 3 Number of studies by specific method used and impact type. (a)** Environmental studies. **(b)** Well-being studies. The vertical axis of both panels shows the specific method used, while the horizontal axis shows the impact type considered

## 3.7 AI research frames environmental impacts positively, but shows more concern for well-being

Classifying studies by overall sentiment on AI impact (positive, negative, or neutral) reveals a striking asymmetry between the two research streams. Environmental studies are much more optimistic about the impacts of AI than well-being studies. Overall, 82% of environmental studies suggest positive impacts, while only 8% discuss negative effects. This finding contrasts sharply with research on human well-being, where positions are more evenly split: 46% portray AI as detrimental and 44% portray it as beneficial (Fig. 4a).

Such divergence may reflect differences in disciplinary focus, scope, analytical level, scale, and time horizon, as well as differences in worldviews (e.g. optimistic vs. pessimistic) and normative assumptions between scholars of different disciplines. Well-being studies, which are often grounded in sociology and critical social sciences, usually examine broad, longer-term societal impacts. Environmental studies, which have so far predominantly originated in engineering and computer science, tend to focus on immediate outcomes at the micro scale, such as efficiency gains and resource optimisation. Furthermore, environmental studies frequently focus on applications designed to improve environmental performance while giving limited consideration to upstream supply-chain and systemic impacts, suggesting that selection effects and narrow system boundaries may also help explain the observed pattern.

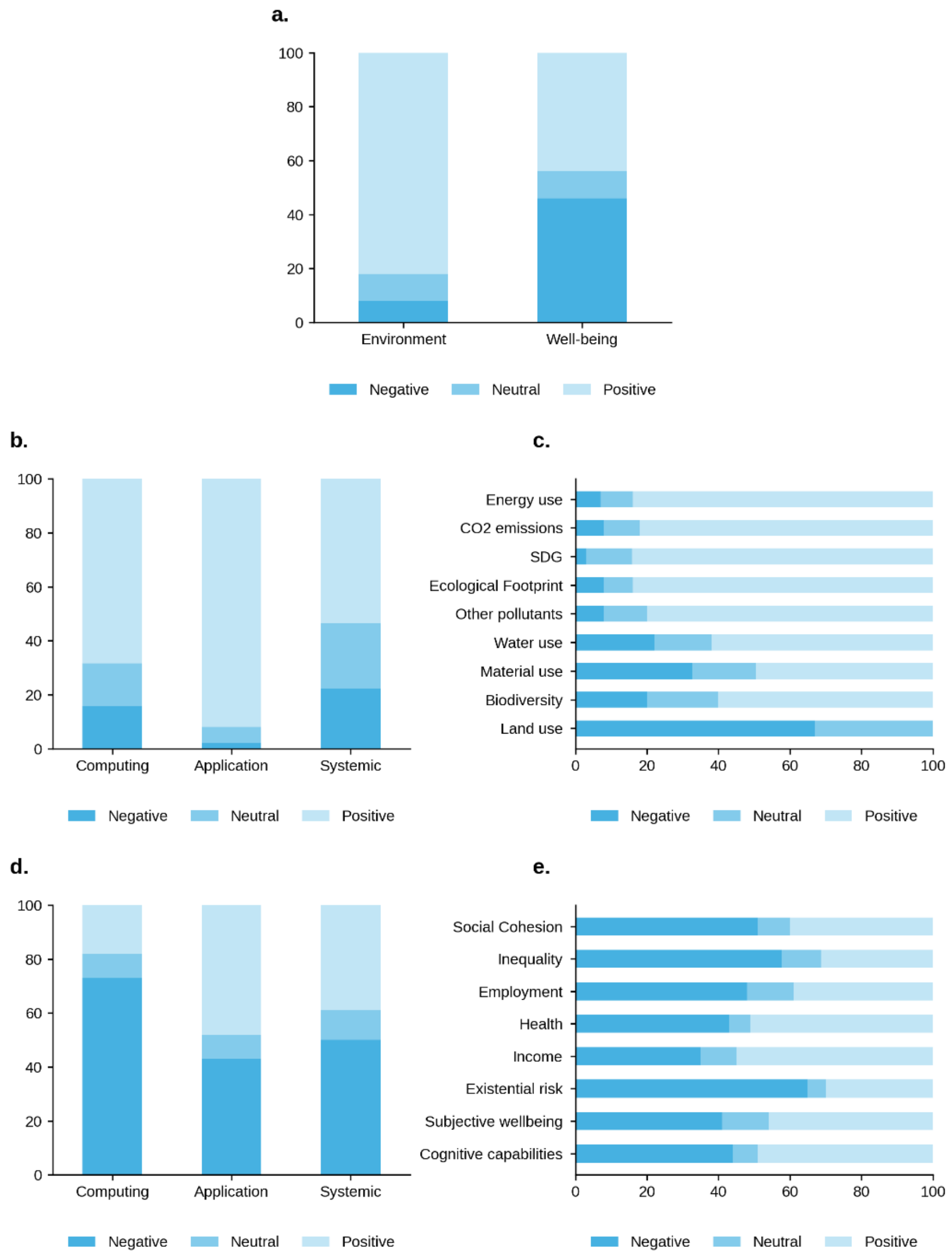

**Fig. 4 Percentage of studies by overall sentiment. (a)** Percentage of environmental and well-being studies by overall sentiment on whether AI's impacts are predominantly positive, neutral, or negative. **(b)** Percentage of environmental studies by overall sentiment grouped by the level at which impacts arise (computing-related, application-level, or systemic). **(c)** Percentage of environmental studies by impact type and overall sentiment. **(d)** Percentage of well-being studies by overall sentiment grouped by the level at which impacts arise (computing-related, application-level, or systemic). **(e)** Percentage of well-being studies by impact type and overall sentiment

### 3.8 Application-level studies drive environmental optimism

The predominance of application-level studies helps explain the overall positive framing of environmental research, as 92% of application-level studies have a positive sentiment (Fig. 4b). By comparison, 69% of computing-related studies adopt a positive perspective, while only 53% of systemic studies are positive. While computing-related studies usually consider the environmental cost of making AI (e.g. energy used for training), they also highlight efficiency improvements, contributing to the positive framing. Application-level studies, by contrast, tend to present technologies intended to improve environmental performance (e.g. smart grids). At the systemic level, studies often focus on rebound effects and other impacts resulting from changes in consumer and institutional behaviour that typically increase consumption, and with it environmental pressure. Studies considering less-explored indicators, such as water use, material use, biodiversity, and — in particular — land use, also tend to be more negative in their sentiment (Fig. 4c).

### 3.9 Sentiment is positive for some well-being impacts, but negative for others

In the well-being literature, sentiment varies significantly by impact type. Impacts on income and health are predominantly expected to be positive (55% and 51% respectively; Fig. 4e). Studies on cognitive impacts and subjective well-being are slightly more positive than negative (49% positive compared with 44% negative for cognition, and 46% versus 41% for subjective well-being), likely because many are intervention studies testing tools intended to improve well-being (e.g. emotional support tools). By contrast, the areas with the most negative sentiment are existential risk (65% negative), inequality (57%), social cohesion (51%), and employment (48%). Some studies considering existential risks still express positive sentiment, either because they view these risks as manageable via alignment or policy, or because they emphasise other benefits. Considering the level at which impacts arise, computing-related studies, which assess upstream supply chain impacts, tend to have a predominantly negative sentiment (73%; Fig. 4d).

## 4. Qualitative results

Within the 1,291 studies surveyed, concern regarding the environmental impact of AI is largely driven by the growing scale of LLMs. Application-level studies suggest that AI may improve firms' environmental performance. However, systemic-level studies suggest that efficiency gains may generate rebound effects or even backfire, targeted advertising may increase consumption, and AI uptake might lock in unsustainable socio-technical configurations. On the well-being side, effects on growth and employment remain highly uncertain, while AI is expected to worsen inequality and social cohesion through concentration of power, social exclusion, misinformation, and surveillance. Substantial benefits are expected for human health, but concerns are growing that sustained LLM use may increase cognitive offloading, anxiety, and social isolation, affecting cognitive capabilities and subjective well-being. Tables 1 and 2 summarise the key findings and research gaps for environmental and well-being impacts, respectively.

### 4.1 The explosive growth of LLMs and computing-related impacts drive environmental concerns

Compute requirements for data-centric AI have surged over the past decade, driving escalating resource demands (Xiao et al. 2025). Improvements in AI are still led by increasing model size, despite recent evidence of diminishing accuracy gains (Varoquaux et al. 2025).

Although most studies focus on energy use and $CO_2$ emissions from training, some assessments note that inference emissions often exceed training emissions, in some cases by an order of magnitude (Chien et al. 2023). Data centres are often located in regions with above-average carbon intensity (Guidi et al. 2024) and water stress (Li et al. 2025), competing with local communities for electricity and water and causing environmental conflicts (Lehuedé 2025).

Beyond operations, embodied emissions from hardware and infrastructure are increasingly being considered, but remain uncertain, with many studies placing them around 30–50% (Wu et al. 2022). Semiconductor fabrication facilities withdraw millions of litres of water daily and release toxic by-products (Frost and Hua 2019), while mining for critical minerals and e-waste disposal severely affect ecosystems and marginalised communities (Lehuedé 2025; Sloane et al. 2025; Wang et al. 2024b).

Projections indicate that global AI demand could drive 4.2–6.6 billion $m^3$ of water withdrawals by 2027 (Li et al. 2025) and 200–400 TWh/year of electricity consumption by 2030 (Kamiya and Coroamă 2025), yet transparency remains very limited. A recent analysis found that 84% of the use of leading LLMs involved models with no environmental disclosure (Luccioni et al. 2025c).

**Table 1 Key findings and research gaps on AI's environmental impacts.** The table reports key findings and underexplored topics organised by the level at which impacts arise. Findings were selected based on impact severity and degree of confidence. See Supplementary Table 1 for the complete list, with references

| Environmental sustainability impacts | | |
|---|---|---|
| **Level** | **Key findings** | **Research gaps** |
| **Computing-related impacts** | • Escalating computing requirements are driving increasing energy use and $CO_2$ emissions for diminishing accuracy gains<br>• Inference emissions may exceed training emissions at deployment scale<br>• Embodied emissions from hardware and infrastructure may account for ~30–50% of the life-cycle carbon footprint<br>• Training a single LLM can use over 5 million litres of freshwater, while inference can require ~500 ml for each 10–50 responses<br>• Data centres are energy- and water-intensive and are often sited on carbon-intensive grids or in water-stressed regions<br>• Semiconductor fabrication uses millions of litres of ultrapure water daily and generates hazardous waste<br>• Mining for AI-critical materials such as lithium, cobalt, and rare earths causes water depletion, pollution, and biodiversity loss<br>• AI accelerates hardware turnover, contributing to global e-waste | • Water, material, land use, non-carbon pollutants (including e-waste and toxics), and biodiversity impacts<br>• Inference, infrastructure and hardware manufacturing, mining, trade, and end-of-life disposal phases<br>• Regional, community-level impacts<br>• Firm-level standardised disclosures<br>• Environmental sustainability metrics in benchmarking and model evaluation |
| **Application-level impacts** | • AI can improve firms' energy and resource efficiency<br>• Environmental benefits vary by sector, region, firm type, and deployment stage, with stronger effects in high-income, industry-dense and highly regulated areas, and in sectors like energy, manufacturing, agriculture, and smart cities<br>• AI supports environmental monitoring | • Causal pathways<br>• AI's role in carbon-intensive sectors<br>• Comparison of green and brown AI impacts<br>• Impacts beyond high-income countries and China |
| **Systemic impacts** | • Efficiency gains in AI are triggering the rebound effect: efficiency improves, yet aggregate electricity use continues to grow<br>• AI-driven efficiency gains in other sectors may induce sectoral rebound<br>• AI-driven personalisation and advertising may intensify consumption<br>• AI could enhance long-term climate planning and decision-making, supporting climate-change mitigation<br>• AI might entrench lock-in and path dependencies, as market incentives favour scalable, high-profit uses over sustainability-oriented ones | • Empirical studies<br>• Quantification of rebound effects<br>• Path dependencies and lock-in<br>• Methods to assess other systemic effects |

### 4.2 AI may improve firms' environmental performance

Application-level studies frequently report improvements in firms' environmental performance (42%). Their findings point to nonlinear effects varying by region, sector, and firm type (Chen et al. 2024; Fu et al. 2024; Shang et al. 2024; Zhang et al. 2024). Improvements are mainly observed in high-income, industry-dense, and highly regulated contexts, and in sectors such as energy, manufacturing, agriculture, smart cities, and environmental monitoring (Li et al. 2024; Liu et al. 2022; Wang et al. 2024a). However, many studies rely on indirect proxies for AI adoption and impose linear assumptions, leaving the causal mechanisms connecting AI to environmental performance poorly understood.

Despite the focus on application-level impacts, we identified no studies that explicitly assess whether AI may reinforce emissions-intensive sectors (e.g. fossil fuels, personalised consumption), highlighting a critical evidence gap.

### 4.3 AI may cause rebound effects, increase consumption, and delay structural reforms

Environmental research on systemic impacts remains limited and predominantly conceptual. A common argument is that efficiency improvements alone do not guarantee net environmental benefits (Wright et al. 2025) and may trigger rebound effects that can offset, or even reverse, expected environmental benefits (Gaffney et al. 2025; Luccioni et al. 2025b). Within AI systems, lower computational costs may lead to wider deployment, more frequent use, investment in larger models, or resource-intensive hardware upgrades. Where AI is used to improve efficiency in sectors such as transport or manufacturing, increased activity may similarly offset or reverse the expected environmental benefits.

Across the reviewed literature, 31% of systemic-impact studies address rebound dynamics, while 40% examine AI-led shifts in consumer behaviour. Key examples include personalisation, recommender systems, and targeted advertising, which may stimulate consumption by increasing the relevance and frequency of commercial content and reducing barriers to purchasing, thereby increasing associated environmental pressures.

Other systemic impacts acknowledged in the literature include the risk of reinforcing existing socio-technical configurations associated with high energy and resource demands. For example, the development of autonomous vehicles might reinforce private-car-oriented transport at the expense of lower-carbon alternatives (Kaack et al. 2022; Wright et al. 2025). Although AI may also support long-term climate planning and mitigation, market incentives may prioritise scalable, high-profit applications, including advertising and fossil-fuel optimisation, over sustainability-oriented ones (Rikap 2024; Varoquaux et al. 2025). AI may therefore improve the efficiency of existing systems while reinforcing path dependencies and delaying more fundamental structural change (Kaack et al. 2022; Varoquaux et al. 2025; Wright et al. 2025).

### 4.4 AI's impacts on employment and economic growth remain uncertain, but inequality is likely to increase

Employment research highlights both productivity gains and job displacement as potential impacts of AI (Acemoglu 2024; Acemoglu and Restrepo 2019; Ozgul et al. 2024; Qiao et al. 2026). Higher-wage cognitive tasks, such as writing and coding, appear particularly exposed, while relational, scientific, and critical-thinking roles remain relatively resilient (Eloundou et al. 2024; Felten et al. 2021). Senior positions tend to retain a human premium, whereas junior and entry-level roles face greater automation risk (Xie and Avila 2025). Experimental evidence also suggests that productivity gains may be greatest among novice and lower-performing workers, potentially narrowing performance gaps (Noy and Zhang 2023).

While AI might replace humans in an expanding range of tasks, its supply chains rely extensively on on-demand work and microwork to perform tasks such as data collection, annotation, content moderation, and model evaluation. This work is often precarious, low-paid, and disproportionately undertaken by women and workers in the Global South (Altenried 2020; Mohla et al. 2024; Nwachukwu et al. 2023; Vargas 2025). Yet, effects on job quality, including autonomy, satisfaction, and meaning, remain poorly understood. Where AI is used to monitor workers and standardise work, it may reduce autonomy and increase psychosocial stress (Deranty and Corbin 2022).

Current evidence on productivity gains shows modest increases concentrated in data-intensive sectors. Economic growth projections span the full spectrum, from negligible to potentially unbounded, depending on assumptions (Trammell and Korinek 2023; Wynne and Derr 2025). This wide range primarily reflects assumptions about the pace of AI capability advancement and the extent of economic transformation, from gradual productivity improvements to widespread automation of production and research.

Distributionally, most studies find that AI is likely to shift income from labour to capital, weakening worker bargaining power and raising wealth-concentration risk (Acemoglu and Restrepo 2018; Acemoglu et al. 2022; Autor and Salomons 2018; Hubmer and Restrepo 2021). Data-ownership asymmetries provide a further mechanism through which gains may become concentrated, as firms can extract and monetise data and creative or intellectual works without compensating their original creators (Epstein et al. 2023; Mulligan and Godsiff 2023; Precel et al. 2024).

AI may expand access to services, but these benefits are not necessarily distributed equitably. Biases in training data and existing social inequalities may be reproduced or amplified by AI systems, resulting in discriminatory outcomes across gender, sexual orientation, ethnicity, age, and disability status, including in access to work and essential services (An et al. 2025; Capraro et al. 2024; Hofmann et al. 2024; Neumann et al. 2024).

Additionally, the current AI development paradigm risks exacerbating global asymmetries in technological capacity and in the distribution of burdens and benefits (Hagerty and Rubinov 2019; Mirza et al. 2024; Mohamed et al. 2020; Vargas 2025). Unequal access to connectivity, computing resources, and AI literacy may limit the ability of lower-income countries and marginalised communities to shape and benefit from AI. Meanwhile, AI supply chains may extract data, labour, and environmental resources from these communities without commensurate compensation, concentrating technological and economic gains elsewhere. Despite these concerns, research on intersectional fairness (Ovalle et al. 2023), wealth concentration (Hussein et al. 2021), and global asymmetries (Abungu et al. 2023; Arora 2024; Sengupta et al. 2023) remains limited.

**Table 2 Key findings and research gaps on AI's impacts on human well-being.** The table reports key findings and underexplored topics organised by impact type. Findings were selected based on impact severity and degree of confidence. See Supplementary Table 2 for the complete list, with references

| Human well-being impacts | | |
|---|---|---|
| **Impact type** | **Key findings** | **Research gaps** |
| **Social cohesion** | • AI can support consensus-building in large-scale decisions<br>• AI may increase concentration of power among a few players<br>• AI risks intensifying surveillance and amplifying misinformation<br>• AI may overlook minority positions and contributions<br>• AI might weaken social trust and community bonds | • Empirical evidence<br>• Long-term impact on social relationships |
| **Inequality** | • AI may amplify social biases and discrimination<br>• Privacy and ownership asymmetries may exacerbate inequality<br>• Unequal access and benefits in AI may amplify global asymmetries | • Intersectional fairness<br>• Global North–South divide in AI<br>• Impact on wealth concentration |
| **Employment** | • AI adoption drives both displacement and productivity gains<br>• AI contributes to declining labour share<br>• AI supply chains rely on precarious gig work and microwork | • Substitution–complementarity dynamics<br>• Impacts on job quality and informal work<br>• Global microwork impacts |
| **Health** | • AI improves diagnosis, treatment, and surgery<br>• AI in health is most effective as a complement to human care<br>• AI may amplify health inequities | • Empirical evidence<br>• Informed consent and liability frameworks |
| **Economic growth and income** | • AI adoption yields modest firm-level productivity gains, concentrated in data-intensive sectors<br>• AI may increase GDP growth, with estimates from modest to very large | • AI modelling/measurement ambiguity<br>• Empirical validation of macro models<br>• Bounding GDP growth estimates |
| **Existential risk** | • Jailbreaks and misuse may increase bioweapon and cyberattack risks<br>• Autonomous weapons proliferation may escalate conflicts<br>• Emergence of superintelligence might displace human agency | • Alignment strategies<br>• Probabilities and early warning signs |
| **Subjective well-being** | • AI broadens access to mental-health and emotional support<br>• AI has lowered current and expected subjective well-being<br>• AI may increase or decrease workers' well-being<br>• Widespread AI use may increase risks of overdependence and isolation | • Empirical research<br>• Multi-dimensional well-being |
| **Cognitive capabilities** | • AI-mediated personalised learning may improve educational outcomes<br>• LLM exposure may produce cognitive offloading, memory weakening, and homogenisation of thought and language | • Empirical research<br>• Long-term effects (e.g. cognitive debt, decreased autonomy) |

### 4.5 Arms-race dynamic increases existential and social-cohesion risks

The literature suggests that AI can support consensus-building in large-scale decisions and participatory governance (McKinney 2024; Tessler et al. 2024), yet it may also amplify misinformation, weaken social trust, increase surveillance, and concentrate power (Kalluri et al. 2023; Mitra et al. 2025; Shoaib et al. 2023; Vargas 2025; Verdegem 2024). Deepfakes and other synthetic media may compromise evidence authentication, public understanding, and electoral integrity (Shoaib et al. 2023), while facilitating the creation and dissemination of non-consensual and violent content (Birhane 2022; Shoaib et al. 2023). AI-based profiling and political microtargeting could further distort public discourse, fuel misinformation, and erode trust in institutions (Mitra et al. 2025; Pappalardo et al. 2024).

AI may also intensify surveillance through facial recognition, video analytics, predictive policing, and other systems that extract and analyse data about individuals, potentially reinforcing discriminatory monitoring and compromising privacy, civic freedoms, and democratic participation (Hussein et al. 2021; Kalluri et al. 2023; Vargas 2025). Although AI can democratise access to knowledge, it may centralise its provision and "average away" under-represented contributions, risking marginalisation and loss of cultural diversity (Bender et al. 2021; Mohamed et al. 2020).

Rising compute and capital thresholds narrow university and small-organisation participation in frontier AI, concentrating capability in a few large firms (Pachegowda 2023). This concentration of corporate power is increasingly intertwined with military power through states' dependence on privately controlled infrastructure and dual-use technologies, the expansion of defence-industry partnerships, and a growing alignment of corporate and military interests (Coveri et al. 2025). AI is already being deployed in warfare for surveillance, targeting, and autonomous operations, potentially increasing the scale and speed of conflict and reducing human control over targeting and the use of force (Coveri et al. 2025; Tamburrini 2024).

Existential concerns focus on misaligned superintelligence, autonomous weapons proliferation, and misuse scenarios such as cyberattacks or bioweapon development (Goldfarb 2024; Tamburrini 2024; Turchin and Denkenberger 2020). These risks are amplified by an arms race dynamic, with governments and companies compelled to build ever larger models to pre-empt rivals for the sake of national security. Under these competitive pressures, research on AI safety receives substantially less funding than the development of AI capabilities (Han et al. 2020; Varoquaux et al. 2025).

### 4.6 AI can improve healthcare, but may reduce subjective well-being and cognitive capabilities

There is an extensive literature on the impacts of AI on health. Many studies demonstrate the benefits of AI for diagnosis, treatment, and surgery, especially when it complements rather than replaces human expertise (Saccà et al. 2023). However, these benefits may be distributed unequally because of biased systems and unequal access (Paik et al. 2023; Rinderknecht et al. 2024; Van Kolfschooten 2023). Gaps in informed consent, clinical oversight, privacy, and liability, together with over-reliance on conversational agents, also raise concerns (Mohammad Amini et al. 2023; Rosic 2024; Saadat et al. 2024).

Impacts on subjective well-being and cognition are far less studied. AI-based mental-health applications and conversational agents may expand access to psychological support and emotional companionship, but may also foster overdependence, reduce autonomy, and displace opportunities for human connection (Alabed et al. 2024). At work, AI may improve workers' well-being by reducing repetitive work or supporting workers' capabilities, or diminish it by increasing surveillance, work intensity, insecurity, or loss of autonomy (Peppiatt 2024). The first empirical study on the impacts of AI on subjective well-being, spanning 137 countries, finds a negative association between AI adoption and both current and expected subjective well-being, likely because AI adoption increases anxiety about job security, income stability, and privacy (Zhao et al. 2024).

Concern about AI's cognitive effects has recently emerged, driven by extensive use of LLMs among young people and in education. Although personalised learning tools may improve educational access and performance, growing reliance on LLMs raises questions about how students engage with and retain knowledge (Qasmi and Fatima 2024). Preliminary evidence shows that prolonged exposure to LLMs may lower mental engagement, impair memory retention, and homogenise language and thought (Dergaa et al. 2024; Gerlich 2025; Kosmyna et al. 2025). These findings raise concerns about cognitive offloading and the possible accumulation of cognitive debt, whereby short-term improvements in speed or performance reduce opportunities to practise critical thinking, problem-solving, and creativity.

More empirical and longitudinal research is needed to assess AI's impacts on human well-being, including mental health, cognition, life satisfaction, and time use.

## 5. A unified framework for AI's environmental and well-being impacts

Taken together, our findings suggest that impacts on the environment and well-being cannot be regarded as separate consequences of AI. They may stem from the same supply chains, infrastructure, applications, and institutional changes, while benefits in one area may coexist with costs in another. Assessing these impacts requires a common analytical structure that considers both dimensions across the different levels at which impacts arise.

In an influential article, Kaack et al. (2022) classify the climate impacts of AI into three categories (or levels): computing-related, application-level, and systemic impacts. We argue that this classification could also be interpreted more broadly to cover all impacts of AI on environmental sustainability, not just those relating to greenhouse gas emissions. Furthermore, we argue that the social and well-being impacts of AI also span these three levels.

We therefore extend the framework in two respects. First, we apply it to environmental impacts beyond greenhouse gas emissions, including water, materials, land, biodiversity, and other pollutants. Second, we apply the same three categories to the impacts of AI on human well-being. Well-being impacts may originate in upstream supply chains (e.g. precarious microwork, unequal exchange), arise from specific applications (e.g. labour displacement, misinformation), or result from systemic shifts in behaviour and institutions (e.g. power concentration, widening digital divide). Therefore, building on our empirical results, we propose a unified framework that enables the simultaneous consideration of both environmental and human well-being impacts at the computing, application, and systemic levels (Fig. 5).

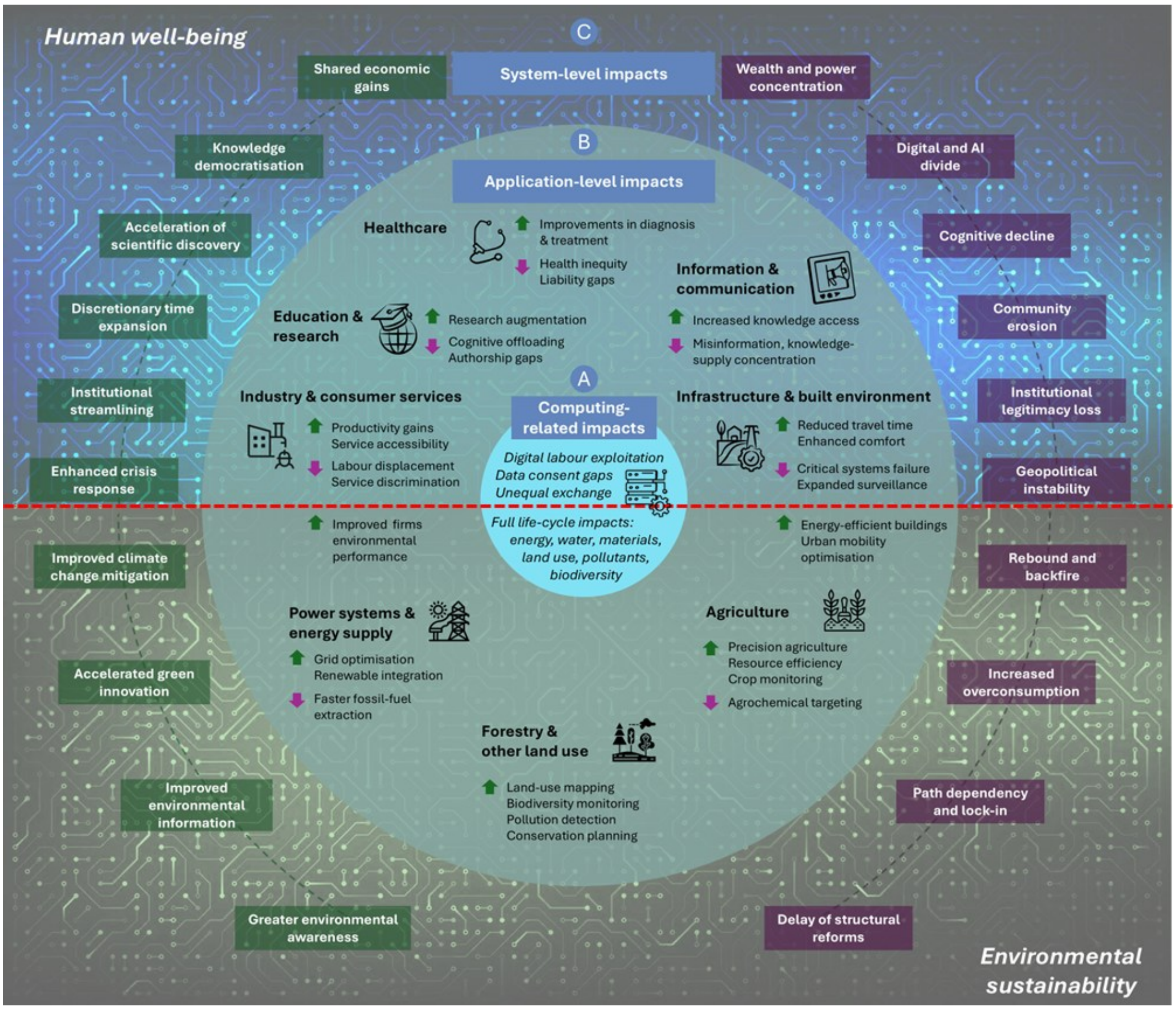

**Fig. 5 Unified framework for assessing the environmental and well-being impacts of AI.** Adapted from the framework initially proposed by Kaack et al. (2022), which classified AI climate impacts into three categories: A (computing-related), B (application-level), and C (systemic). We extend this structure in two ways: first, we include environmental sustainability impacts beyond greenhouse gas emissions; second, we integrate human well-being impacts across the same three categories. The upper half of the figure shows impacts on human well-being, while the lower half shows impacts on environmental sustainability. At the application level, upward green arrows indicate likely positive impacts, while downward magenta arrows indicate likely negative impacts. At the systemic level, likely positive impacts are shown in green and positioned on the left, while likely negative impacts are shown in magenta and positioned on the right

The framework places environmental and well-being impacts within a common structure, highlighting how they may stem from the same underlying processes, co-occur, and compound one another. At the computing level, mining, manufacturing of hardware and infrastructure, and end-of-life disposal generate environmental pressures, while simultaneously affecting workers and communities. At the application level, AI deployment in agriculture may affect water, energy, and pesticide use, while also influencing livelihoods and food security. At the systemic level, the growing computational requirements of frontier AI models increase energy and material demands while simultaneously raising barriers to participation and concentrating technological and economic power.

The framework serves four analytical purposes. First, it makes the system boundaries of AI impact assessment explicit by distinguishing computing-related, application-level, and systemic impacts. This helps prevent the immediate effects of specific applications from being treated as a complete assessment of AI. Second, by applying the same three-level framework to both environmental sustainability and human well-being impacts, it provides a common basis for systematically comparing the two bodies of literature. Third, it provides a systematic means of identifying asymmetries, blind spots, and research gaps (including the limited attention paid to systemic environmental effects and computing-related well-being impacts). Fourth, the framework makes interactions visible both between environmental sustainability and human well-being and across the different levels at which impacts arise, including potential trade-offs.

As an example of a trade-off between environmental sustainability and human well-being, AI could increase industrial resource efficiency while displacing workers, or expand discretionary time while driving higher consumption and rebound effects. Trade-offs may also arise within the same dimension but across different levels of the framework. For example, improvements in computational efficiency could reduce the resources required for a single application but lead to wider adoption and increase overall resource use. Therefore, the direction and extent of AI's impacts depend on the interactions between the different dimensions (environmental sustainability, human well-being) and the different levels at which the impacts arise (computing-related, application, systemic). The framework makes these interactions explicit.

Furthermore, the framework can help relate impacts to different policy levers according to the levels at which they arise. Computing-related impacts may be addressed by enforcing environmental and labour standards within data and AI supply chains. Application-level impacts require domain-specific instruments, such as clinical validation and informed-consent requirements for AI in healthcare, and assessments of emissions, induced travel, and modal shifts for AI applications in transport. Systemic impacts, on the other hand, will need broader policies, regulations, and international agreements in order to be tackled. Examples include emissions and resource caps for AI infrastructure, and international agreements prohibiting autonomous weapons.

By considering interconnected impacts across both dimensions and all three levels, the unified framework provides a basis for assessing whether AI development contributes to environmental sustainability and human flourishing.

## 6. Discussion and future research

Overall, our systematic review reveals a divided literature. The environmental literature is dominated by application-level studies, which largely focus on how AI applications could reduce energy use and $CO_2$ emissions. However, these studies tend to overlook computing-related costs and neglect important systemic effects. The well-being literature takes a broader and more macro view, but it is also more theoretical. Unlike the environmental literature, where sentiment is markedly positive, sentiment on AI's well-being impacts is roughly split between positive and negative, but this split masks variations across dimensions. While AI is expected to have a positive impact on income and health, it carries existential risks and is expected to have a negative impact on social cohesion, inequality, and employment.

The divergence in sentiment between the two fields partly reflects differences in disciplinary framing, the impacts considered, the methods used, the analytical scales adopted, and the levels at which impacts are examined. The current literature therefore does not provide a neutral overview of AI's impacts, but reflects the questions, methods, and system boundaries through which those impacts have been assessed.

These findings should be interpreted in light of several limitations. First, the search strings and databases that we used may not capture the full range of relevant work across disciplines. We included studies from arXiv and NBER Working Papers to maximise disciplinary breadth and recency, but these databases include preprints and working papers that may not have been peer-reviewed. We mitigated this limitation by conducting quality appraisal that was blind to authorship and publication source. Second, industry influence might have shaped some included studies. Third, relevance-based screening and classification by overall sentiment inevitably involve some subjective judgement, despite efforts to establish a standardised process with objective criteria. Fourth, the use of Microsoft 365 Copilot Enterprise to support data extraction from less-relevant papers may have introduced inaccuracies or biases, although every field and entry was manually checked against the original source. Finally, the AI evidence base is evolving rapidly, and findings may change as models, hardware, and deployment contexts evolve.

Notwithstanding these limitations, our study makes several important contributions. First, to our knowledge, it provides the most comprehensive synthesis to date of AI's environmental and well-being impacts, systematically mapping 1,291 studies across environmental sciences, social sciences, computer science, engineering, medicine, philosophy, and economics, and considering underexplored topics such as biodiversity, subjective well-being, and cognition. Second, by examining environmental and well-being research together, our analysis provides the first systematic comparison of the two literatures and identifies major asymmetries in scope, analytical scale, methods, levels at which impacts are examined, and sentiment. Third, it demonstrates how these disciplinary and analytical choices shape the impacts represented in the evidence base and help explain the predominantly positive framing of environmental impacts relative to the more divided well-being literature.

Fourth, we extend Kaack et al.'s (2022) classification beyond greenhouse gas emissions to other environmental impacts and systematically apply it to human well-being, showing that systemic environmental impacts are critically understudied, while upstream supply-chain social impacts remain largely overlooked. Fifth, building on this analysis, we develop a unified framework that integrates environmental sustainability and human well-being impacts across computing-related, application, and systemic levels, enabling direct comparison and making interactions and trade-offs across dimensions and levels explicit. Finally, the review identifies several major research gaps, providing an empirical basis to help prioritise future work.

A central implication of our review is that, given the pervasiveness of the technology, AI impacts should not be considered in isolation. Environmental, social, and well-being impacts must be considered together. Environmental impacts often have distributional consequences, while social impacts, such as rising inequality, may undermine environmental mitigation (Dong et al. 2024). Similarly, even if AI is deployed for social benefit, it may still have environmental costs. The growing computational requirements of frontier AI, for example, simultaneously increase environmental demands, raise barriers to participation, and concentrate technological and economic power. Assessment must therefore consider how impacts interact across dimensions and levels, who receives the benefits, who bears the burdens, and which technical, economic, social, and political arrangements produce these impacts.

Based on our findings, we identify four priorities for future research. First, we argue that future research on AI's environmental impacts must broaden its scope beyond energy use and $CO_2$ emissions to include water, materials, land use, non-carbon pollutants, and biodiversity. These dimensions are critical because AI depends on resource-intensive infrastructure and hardware, relies on extractive global supply chains, and contributes to escalating e-waste. Addressing these issues requires not only life-cycle approaches, but also analysis of global supply chains, for example using input–output methods (Kitzes 2013), which allow environmental burdens to be quantified as they propagate across interdependent sectors and countries.

Second, research should devote greater attention to systemic environmental impacts. The literature largely frames AI's environmental impacts as positive, in part because existing studies focus more on how AI can solve environmental problems than on assessing its environmental costs or broader systemic impacts. A broader perspective is needed to comprehensively assess the positive and negative impacts of AI, especially in the context of environmental limits (O'Neill and Creutzig 2026), while jointly considering computing-related, application-level, and systemic impacts. At the application level, research should also assess applications that may increase environmental pressures, rather than focusing predominantly on those designed to improve environmental performance. At the systemic level, future work should quantify rebound effects under different development and deployment scenarios.

Third, social and well-being impacts deserve more empirical investigation. Attention is growing regarding the risks that AI poses to social cohesion (Creutzig et al. 2026). Although the potential impacts are serious, from threats to democracy and social trust to the imposition of Western culture and geopolitical risks, empirical evidence remains scarce. In healthcare, studies should examine real-world clinical and equity outcomes rather than relying predominantly on technical performance. Employment research should distinguish occupational exposure from realised displacement and determine under which conditions AI complements or substitutes for workers. Beyond employment levels, research should devote greater attention to job quality, autonomy, meaning, and the subjective dimensions of work. Research on subjective well-being and cognition, which remain particularly understudied, should examine whether and how AI affects these dimensions over time, given emerging concerns that AI may reduce current and expected subjective well-being (Zhao et al. 2024) and that sustained reliance on LLMs may foster cognitive offloading and homogenisation of thought (Dergaa et al. 2024; Gerlich 2025; Kosmyna et al. 2025).

Fourth, future research should investigate social impacts arising from AI's upstream supply chains, including data consent gaps, community burdens from infrastructure siting, unequal exchange, and precarious data work. Qualitative approaches, particularly participatory methods, can support these investigations by involving affected workers and communities in identifying relevant impacts and providing contextual evidence.

Together, these priorities would support a more comprehensive assessment of AI as a socio-technical system. Understanding the scope, scale, and direction of AI's impacts requires integrating environmental and well-being dimensions, examining interactions across computing-related, application, and systemic levels, and considering how benefits and burdens are distributed.

## 7. Conclusion

To examine and integrate research on AI's impacts on environmental sustainability and human well-being, we conducted a systematic literature review of 1,291 studies, including 523 environmental studies and 768 human well-being studies. We find that the environmental literature has largely focused on energy use and $CO_2$ emissions. Environmental studies tend to apply quantitative approaches at the micro scale, and while sentiment is markedly positive, systemic effects are largely overlooked. By contrast, the well-being literature is predominantly conceptual and adopts a more macro perspective. Over the full well-being literature, sentiment is roughly split between positive and negative impacts, but at the level of specific well-being variables it diverges. AI is expected to have a positive impact on income and health, but a negative impact on social cohesion, inequality, and employment. The literature rarely addresses cognition, subjective well-being, or impacts arising from upstream supply chains.

We argue that environmental assessments should account for the entire AI life cycle and the full range of environmental impacts, including water, materials, biodiversity, land use, and non-carbon pollutants. Life-cycle assessment and environmentally extended input-output analysis can support these assessments, with the latter tracking global supply chains and burden shifts. More focus also needs to be placed on systemic impacts, including estimating rebound effects through scenario-based approaches.

For social and well-being impacts, more empirical research is needed, including qualitative and participatory approaches. Research should examine real-world outcomes in healthcare and employment, including effects on job quality, and strengthen the evidence on social cohesion, subjective well-being, and cognition. Greater attention should also be paid to impacts arising from upstream supply chains, such as data-worker conditions, local community impacts, and ecologically unequal exchange.

To support this research agenda, we extend Kaack et al.'s (2022) classification beyond greenhouse gas emissions to other environmental impacts and systematically apply it to impacts on human well-being. The resulting unified framework provides a common structure for tracing environmental and well-being impacts across computing-related, application, and systemic levels, highlighting interactions and trade-offs and promoting a more comprehensive assessment of AI impacts.

AI is a socio-technical system shaped by social choices. To guide policy towards mitigating risks while realising benefits, we need a comprehensive approach that traces impacts and the technical, economic, and political drivers behind them. The literature envisages a wide range of futures, from higher economic growth, accelerated scientific discovery, and improved health on the one hand, to mass unemployment, severe societal and environmental disruption, and existential risk on the other. Given this uncertainty and the rapid pace of deployment, gaining a better understanding of AI's potential benefits and risks — and acting on this information in a timely manner — is critical.

## Acknowledgements

This research was funded by the European Union in the framework of the Horizon Europe Research and Innovation Programme under grant agreement number 101137914 (MAPS: "Models, Assessment, and Policies for Sustainability").

## Author contributions

All authors contributed to the conceptual design of the systematic literature review. N.L.C. developed the review protocol, conducted screening, data extraction and analysis, and prepared the initial draft of the article. D.W.O. and T.D. contributed to the refinement of the review protocol and provided ongoing guidance and critical feedback throughout the research process. All authors contributed to the interpretation of the findings and the writing of the article.

## Competing interests

The authors declare no competing interests.

## $CO_2$ emissions statement

Using available estimates of average energy consumption per token for LLM inference, we calculated that the data extraction performed with the assistance of Microsoft 365 Copilot Enterprise for this study resulted in roughly 39 kg of $CO_2$ emissions. Calculation assumptions are reported in the Supplementary Information.

## References

Abungu C, Malonza M, Adan SN (2023) Can apparent bystanders distinctively shape an outcome? Global South countries and global catastrophic risk-focused governance of artificial intelligence. Preprint at https://doi.org/10.48550/arXiv.2312.04616

Acemoglu D (2024) Harms of AI. In: Bullock JB, Chen Y-C, Himmelreich J, et al. (eds) The Oxford Handbook of AI Governance. Oxford University Press, pp 660–706. https://doi.org/10.1093/oxfordhb/9780197579329.013.65

Acemoglu D, Anderson GW, Beede DN, et al (2022) Automation and the Workforce: A Firm-Level View from the 2019 Annual Business Survey. NBER Working Paper 30659. National Bureau of Economic Research. https://doi.org/10.3386/w30659

Acemoglu D, Restrepo P (2018) The Race between Man and Machine: Implications of Technology for Growth, Factor Shares, and Employment. American Economic Review 108:1488–1542. https://doi.org/10.1257/aer.20160696

Acemoglu D, Restrepo P (2019) Artificial Intelligence, Automation, and Work. In: Agrawal A, Gans J, Goldfarb A (eds) The Economics of Artificial Intelligence: An Agenda. University of Chicago Press, pp 197–236. https://doi.org/10.7208/chicago/9780226613475.003.0008

Alabed A, Javornik A, Gregory-Smith D, Casey R (2024) More than just a chat: a taxonomy of consumers' relationships with conversational AI agents and their well-being implications. European Journal of Marketing 58:373–409. https://doi.org/10.1108/EJM-01-2023-0037

Altenried M (2020) The platform as factory: Crowdwork and the hidden labour behind artificial intelligence. Capital & Class 44:145–158. https://doi.org/10.1177/0309816819899410

An J, Huang D, Lin C, Tai M (2025) Measuring gender and racial biases in large language models: Intersectional evidence from automated resume evaluation. PNAS Nexus 4:pgaf089. https://doi.org/10.1093/pnasnexus/pgaf089

Arora P (2024) Creative data justice: a decolonial and indigenous framework to assess creativity and artificial intelligence. Information, Communication & Society 28:2231–2247. https://doi.org/10.1080/1369118X.2024.2420041

Autor D, Salomons A (2018) Is Automation Labor Share-Displacing? Productivity Growth, Employment, and the Labor Share. Brookings Papers on Economic Activity 2018:1–87. https://doi.org/10.1353/eca.2018.0000

Bender EM, Gebru T, McMillan-Major A, Shmitchell S (2021) On the Dangers of Stochastic Parrots: Can Language Models Be Too Big? 🦜. In: Proceedings of the 2021 ACM Conference on Fairness, Accountability, and Transparency. Association for Computing Machinery, New York, NY, USA, pp 610–623. https://doi.org/10.1145/3442188.3445922

Birhane A (2022) Automating Ambiguity: Challenges and Pitfalls of Artificial Intelligence. Preprint at https://doi.org/10.48550/arXiv.2206.04179

Blanchflower DG (2025) Declining Youth Well-being in 167 UN Countries. Does Survey Mode, or Question Matter? NBER Working Paper 33415. National Bureau of Economic Research. https://doi.org/10.3386/w33415

Capraro V, Lentsch A, Acemoglu D, et al (2024) The impact of generative artificial intelligence on socioeconomic inequalities and policy making. PNAS Nexus 3:pgae191. https://doi.org/10.1093/pnasnexus/pgae191

Carmeno NL, Domingos T, O'Neill DW (2026) Systematic Literature Review Protocol: The impacts of artificial intelligence on environmental sustainability and human well-being. protocols.io. https://doi.org/10.17504/protocols.io.j8nlkzjwwl5r/v1

Chancel L, Gómez-Carrera R, Moshrif R, Piketty T (2025) World Inequality Report 2026. World Inequality Lab. https://wir2026.wid.world/

Chen C-T, Khan A, Chen S-C (2024) Modeling the impact of BDA-AI on sustainable innovation ambidexterity and environmental performance. Journal of Big Data 11:124. https://doi.org/10.1186/s40537-024-00995-6

Chien AA, Lin L, Nguyen H, et al (2023) Reducing the Carbon Impact of Generative AI Inference (today and in 2035). In: Proceedings of the 2nd Workshop on Sustainable Computer Systems. ACM, Boston, MA, USA, pp 1–7. https://doi.org/10.1145/3604930.3605705

Chuang C-W, Chang A, Chen M, et al (2022) A Worldwide Bibliometric Analysis of Publications on Artificial Intelligence and Ethics in the Past Seven Decades. Sustainability 14:11125. https://doi.org/10.3390/su141811125

Coveri A, Cozza C, Guarascio D (2025) Big Tech and the US Digital-Military-Industrial Complex. Intereconomics 60:81–87. https://doi.org/10.2478/ie-2025-0017

Creutzig F, Denton F, Hine E, et al (2026) Governing artificial intelligence for planetary health. The Lancet Planetary Health 10:101408. https://doi.org/10.1016/j.lanplh.2025.101408

Deranty J-P, Corbin T (2022) Artificial Intelligence and work: a critical review of recent research from the social sciences. AI & Society 39:675–691. https://doi.org/10.1007/s00146-022-01496-x

Dergaa I, Ben Saad H, Glenn JM, et al (2024) From tools to threats: a reflection on the impact of artificial-intelligence chatbots on cognitive health. Frontiers in Psychology 15:1259845. https://doi.org/10.3389/fpsyg.2024.1259845

Dong Z, Zhang L, Tan C, et al (2024) Does income inequality undermine the carbon abatement benefits of artificial intelligence? Journal of Cleaner Production 472:143437. https://doi.org/10.1016/j.jclepro.2024.143437

Eloundou T, Manning S, Mishkin P, Rock D (2024) GPTs are GPTs: Labor market impact potential of LLMs. Science 384:1306–1308. https://doi.org/10.1126/science.adj0998

Epstein Z, Hertzmann A, Herman L, et al (2023) Art and the science of generative AI: A deeper dive. Science 380:1110–1111. https://doi.org/10.1126/science.adh4451

Felten E, Raj M, Seamans R (2021) Occupational, industry, and geographic exposure to artificial intelligence: A novel dataset and its potential uses. Strategic Management Journal 42:2195–2217. https://doi.org/10.1002/smj.3286

Fraisl D, See L, Fritz S, et al (2025) Leveraging the collaborative power of AI and citizen science for sustainable development. Nature Sustainability 8:125–132. https://doi.org/10.1038/s41893-024-01489-2

Frost K, Hua I (2019) Quantifying spatiotemporal impacts of the interaction of water scarcity and water use by the global semiconductor manufacturing industry. Water Resources and Industry 22:100115. https://doi.org/10.1016/j.wri.2019.100115

Fu Y, Shen Y, Song M, Wang W (2024) Does artificial intelligence reduce corporate energy consumption? New evidence from China. Economic Analysis and Policy 83:548–561. https://doi.org/10.1016/j.eap.2024.07.005

Gaffney O, Luers A, Carrero-Martinez F, et al (2025) The Earth alignment principle for artificial intelligence. Nature Sustainability 8:467–469. https://doi.org/10.1038/s41893-025-01536-6

Gerlich M (2025) AI Tools in Society: Impacts on Cognitive Offloading and the Future of Critical Thinking. Societies 15:6. https://doi.org/10.3390/soc15010006

Goldfarb A (2024) Pause artificial intelligence research? Understanding AI policy challenges. Canadian Journal of Economics 57:363–377. https://doi.org/10.1111/caje.12705

Goldfarb A, Taska B, Teodoridis F (2023) Could machine learning be a general purpose technology? A comparison of emerging technologies using data from online job postings. Research Policy 52:104653. https://doi.org/10.1016/j.respol.2022.104653

Guarda C, Caseiro J, Pires A (2024) Machine learning to enhance sustainable plastics: A review. Journal of Cleaner Production 474:143602. https://doi.org/10.1016/j.jclepro.2024.143602

Guidi G, Dominici F, Gilmour J, et al (2024) Environmental Burden of United States Data Centers in the Artificial Intelligence Era. Preprint at https://doi.org/10.48550/arXiv.2411.09786

Hagendorff T (2024) Mapping the Ethics of Generative AI: A Comprehensive Scoping Review. Minds and Machines 34:39. https://doi.org/10.1007/s11023-024-09694-w

Hagerty A, Rubinov I (2019) Global AI Ethics: A Review of the Social Impacts and Ethical Implications of Artificial Intelligence. Preprint at https://doi.org/10.48550/arXiv.1907.07892

Han TA, Pereira LM, Santos FC, Lenaerts T (2020) To Regulate or Not: A Social Dynamics Analysis of an Idealised AI Race. Journal of Artificial Intelligence Research 69:881–921. https://doi.org/10.1613/jair.1.12225

Hofmann V, Kalluri PR, Jurafsky D, King S (2024) AI generates covertly racist decisions about people based on their dialect. Nature 633:147–154. https://doi.org/10.1038/s41586-024-07856-5

Hubmer J, Restrepo P (2021) Not a Typical Firm: The Joint Dynamics of Firms, Labor Shares, and Capital–Labor Substitution. NBER Working Paper 28579. National Bureau of Economic Research. https://doi.org/10.3386/w28579

Hussein BR, Halimu C, Siddique MT (2021) The Future of Artificial Intelligence and its Social, Economic and Ethical Consequences. Preprint at https://doi.org/10.48550/arXiv.2101.03366

Kaack LH, Donti PL, Strubell E, et al (2022) Aligning artificial intelligence with climate change mitigation. Nature Climate Change 12:518–527. https://doi.org/10.1038/s41558-022-01377-7

Kalluri PR, Agnew W, Cheng M, et al (2023) The Surveillance AI Pipeline. Preprint at https://doi.org/10.48550/arXiv.2309.15084

Kamiya G, Coroamă VC (2025) Data Centre Energy Use: Critical Review of Models and Results. EDNA–IEA 4E TCP. https://www.iea-4e.org/edna/publications/data-centre-energy-use-critical-review-of-models-and-results/

Kitzes J (2013) An Introduction to Environmentally-Extended Input-Output Analysis. Resources 2:489–503. https://doi.org/10.3390/resources2040489

Kosmyna N, Hauptmann E, Yuan YT, et al (2025) Your Brain on ChatGPT: Accumulation of Cognitive Debt when Using an AI Assistant for Essay Writing Task. Preprint at https://doi.org/10.48550/arXiv.2506.08872

Kudina O, van de Poel I (2024) A sociotechnical system perspective on AI. Minds and Machines 34:21. https://doi.org/10.1007/s11023-024-09680-2

Lehuedé S (2025) An elemental ethics for artificial intelligence: water as resistance within AI's value chain. AI & Society 40:1761–1774. https://doi.org/10.1007/s00146-024-01922-2

Li P, Yang J, Islam MA, Ren S (2025) Making AI Less “Thirsty”: Uncovering and Addressing the Secret Water Footprint of AI Models. Communications of the ACM 68:54–63. https://doi.org/10.1145/3724499

Li X, Wang Q, Tang Y (2024) The Impact of Artificial Intelligence Development on Urban Energy Efficiency—Based on the Perspective of Smart City Policy. Sustainability 16:3200. https://doi.org/10.3390/su16083200

Liu J, Qian Y, Yang Y, Yang Z (2022) Can Artificial Intelligence Improve the Energy Efficiency of Manufacturing Companies? Evidence from China. International Journal of Environmental Research and Public Health 19:2091. https://doi.org/10.3390/ijerph19042091

Lu Y, Zhou Y (2021) A review on the economics of artificial intelligence. Journal of Economic Surveys 35:1045–1072. https://doi.org/10.1111/joes.12422

Luccioni AS, Pistilli G, Sefala R, Moorosi N (2025a) Bridging the Gap: Integrating Ethics and Environmental Sustainability in AI Research and Practice. Preprint at https://doi.org/10.48550/arXiv.2504.00797

Luccioni AS, Strubell E, Crawford K (2025b) From Efficiency Gains to Rebound Effects: The Problem of Jevons’ Paradox in AI’s Polarized Environmental Debate. In: Proceedings of the 2025 ACM Conference on Fairness, Accountability, and Transparency. Association for Computing Machinery, New York, NY, USA, pp 76–88. https://doi.org/10.1145/3715275.3732007

Luccioni AS, Gamazaychikov B, Costa TA da, Strubell E (2025c) Misinformation by Omission: The Need for More Environmental Transparency in AI. Preprint at https://doi.org/10.48550/arXiv.2506.15572

Malja SN, Afrasiabi H (2026) Artificial intelligence and society: mapping the research through a systematic review. AI & Society 41:2877–2913. https://doi.org/10.1007/s00146-025-02555-9

Martinez-Millana A, Saez-Saez A, Tornero-Costa R, et al (2022) Artificial intelligence and its impact on the domains of universal health coverage, health emergencies and health promotion: An overview of systematic reviews. International Journal of Medical Informatics 166:104855. https://doi.org/10.1016/j.ijmedinf.2022.104855

McKinney S (2024) Integrating Artificial Intelligence into Citizens’ Assemblies: Benefits, Concerns and Future Pathways. Journal of Deliberative Democracy 20:1556. https://doi.org/10.16997/jdd.1556

Mirza S, Coelho B, Cui Y, et al (2024) Global-Liar: Factuality of LLMs over Time and Geographic Regions. Preprint at https://doi.org/10.48550/arXiv.2401.17839

Mitra B, Cramer H, Gurevich O (2025) Sociotechnical Implications of Generative Artificial Intelligence for Information Access. In: White RW, Shah C (eds) Information Access in the Era of Generative AI. Springer Nature Switzerland, Cham, pp 161–200. https://doi.org/10.1007/978-3-031-73147-1_7

Mohamed S, Png M-T, Isaac W (2020) Decolonial AI: Decolonial Theory as Sociotechnical Foresight in Artificial Intelligence. Philosophy & Technology 33:659–684. https://doi.org/10.1007/s13347-020-00405-8

Mohammad Amini M, Jesus M, Fanaei Sheikholeslami D, et al (2023) Artificial Intelligence Ethics and Challenges in Healthcare Applications: A Comprehensive Review in the Context of the European GDPR Mandate. Machine Learning and Knowledge Extraction 5:1023–1035. https://doi.org/10.3390/make5030053

Mohla S, Bagh B, Guha A (2024) Thinking beyond Bias: Analyzing Multifaceted Impacts and Implications of AI on Gendered Labour. Preprint at https://doi.org/10.48550/arXiv.2406.16207

Mulligan CEA, Godsiff P (2023) Datalism and Data Monopolies in the Era of A.I.: A Research Agenda. Preprint at https://doi.org/10.48550/arXiv.2307.08049

Neumann T, Lee S, De-Arteaga M, et al (2024) Diverse, but Divisive: LLMs Can Exaggerate Gender Differences in Opinion Related to Harms of Misinformation. Preprint at https://doi.org/10.48550/arXiv.2401.16558

Nord M, Angiolillo F, Good God A, Lindberg SI (2025) State of the world 2024: 25 years of autocratization – democracy trumped? Democratization 32:839–864. https://doi.org/10.1080/13510347.2025.2487825

Noy S, Zhang W (2023) Experimental evidence on the productivity effects of generative artificial intelligence. Science 381:187–192. https://doi.org/10.1126/science.adh2586

Nwachukwu NM, Roberts JS, Montoya LN (2023) The Glamorisation of Unpaid Labour: AI and its Influencers. Preprint at https://doi.org/10.48550/arXiv.2308.02399

O'Neill DW, Creutzig F (2026) A strong sustainability approach to AI development. Nature Machine Intelligence 8:642–644. https://doi.org/10.1038/s42256-026-01240-w

Ovalle A, Subramonian A, Gautam V, et al (2023) Factoring the Matrix of Domination: A Critical Review and Reimagination of Intersectionality in AI Fairness. In: Proceedings of the 2023 AAAI/ACM Conference on AI, Ethics, and Society. Association for Computing Machinery, New York, NY, USA, pp 496–511. https://doi.org/10.1145/3600211.3604705

Ozgul P, Fregin M-C, Stops M, et al (2024) High-skilled Human Workers in Non-Routine Jobs are Susceptible to AI Automation but Wage Benefits Differ between Occupations. Preprint at https://doi.org/10.48550/arXiv.2404.06472

Pachegowda C (2023) The Global Impact of AI-Artificial Intelligence: Recent Advances and Future Directions, A Review. Preprint at https://doi.org/10.48550/arXiv.2401.12223

Page MJ, McKenzie JE, Bossuyt PM, et al (2021) The PRISMA 2020 statement: An updated guideline for reporting systematic reviews. PLOS Medicine 18:e1003583. https://doi.org/10.1371/journal.pmed.1003583

Paik KE, Hicklen R, Kaggwa F, et al (2023) Digital Determinants of Health: Health data poverty amplifies existing health disparities—A scoping review. PLOS Digital Health 2:e0000313. https://doi.org/10.1371/journal.pdig.0000313

Pappalardo L, Ferragina E, Citraro S, et al (2024) A survey on the impact of AI-based recommenders on human behaviours: methodologies, outcomes and future directions. Preprint at https://doi.org/10.48550/arXiv.2407.01630

Peng S-Y, Lin C-F, Streinz T (2021) Artificial Intelligence and International Economic Law: A Research and Policy Agenda. In: Peng S-Y, Lin C-F, Streinz T (eds) Artificial Intelligence and International Economic Law: Disruption, Regulation, and Reconfiguration. Cambridge University Press, pp 1–26. https://doi.org/10.1017/9781108954006.002

Peppiatt C (2024) The Future of Work: Inequality, Artificial Intelligence, and What Can Be Done About It. A Literature Review. Preprint at https://doi.org/10.48550/arXiv.2408.13300

Precel H, McDonald A, Hecht B, Vincent N (2024) A Canary in the AI Coal Mine: American Jews May Be Disproportionately Harmed by Intellectual Property Dispossession in Large Language Model Training. In: Proceedings of the CHI Conference on Human Factors in Computing Systems. Association for Computing Machinery, New York, NY, USA, pp 1–17. https://doi.org/10.1145/3613904.3642749

Qasmi F, Fatima SK (2024) Enhancing Education and Well-Being Through Artificial Intelligence: Opportunities and Challenges. Communications in Computer and Information Science 2162 CCIS:154–164. https://doi.org/10.1007/978-3-031-65996-6_13

Qiao D, Rui H, Xiong Q (2026) Artificial Intelligence and Jobs: Has the Inflection Point Arrived? Evidence from an Online Labor Platform. Journal of Management Information Systems 43:5–37. https://doi.org/10.1080/07421222.2025.2602400

Regona M, Yigitcanlar T, Hon C, Teo M (2024) Artificial intelligence and sustainable development goals: Systematic literature review of the construction industry. Sustainable Cities and Society 108:105499. https://doi.org/10.1016/j.scs.2024.105499

Rikap C (2024) Varieties of corporate innovation systems and their interplay with global and national systems: Amazon, Facebook, Google and Microsoft's strategies to produce and appropriate artificial intelligence. Review of International Political Economy 31:1735–1763. https://doi.org/10.1080/09692290.2024.2365757

Rinderknecht F-A, Nwandu L, Daneshjou R, Lester J (2024) The Impact of Artificial Intelligence on Health Equity in Dermatology. Current Dermatology Reports 13:148–155. https://doi.org/10.1007/s13671-024-00436-w

Rosic A (2024) Legal implications of artificial intelligence in health care. Clinics in Dermatology 42:451–459. https://doi.org/10.1016/j.clindermatol.2024.06.014

Saadat A, Siddiqui T, Taseen S, Mughal S (2024) Revolutionising Impacts of Artificial Intelligence on Health Care System and Its Related Medical In-Transparencies. Annals of Biomedical Engineering 52:1546–1548. https://doi.org/10.1007/s10439-023-03343-6

Saccà R, Turrini R, Ausania F, et al (2023) The ménage à trois of healthcare: the actors in after-AI era under patient consent. Frontiers in Medicine 10:1329087. https://doi.org/10.3389/fmed.2023.1329087

Sakschewski B, Caesar L, Andersen L, et al (2025) Planetary Health Check 2025: A Scientific Assessment of the State of the Planet. Potsdam Institute for Climate Impact Research, Potsdam, Germany. https://doi.org/10.48485/pik.2025.017

Sengupta N, Subramanian V, Mukhopadhyay A, Scaria AG (2023) A Global South perspective for ethical algorithms and the State. Nature Machine Intelligence 5:184–186. https://doi.org/10.1038/s42256-023-00621-9

Shang Y, Zhou S, Zhuang D, et al (2024) The impact of artificial intelligence application on enterprise environmental performance: Evidence from microenterprises. Gondwana Research 131:181–195. https://doi.org/10.1016/j.gr.2024.02.012

Shoaib MR, Wang Z, Ahvanooey MT, Zhao J (2023) Deepfakes, Misinformation, and Disinformation in the Era of Frontier AI, Generative AI, and Large AI Models. In: 2023 International Conference on Computer and Applications (ICCA). pp 1–7. https://doi.org/10.1109/ICCA59364.2023.10401723

Sloane M, Moss E, Kennedy S, et al (2025) Materiality and risk in the age of pervasive AI sensors. Nature Machine Intelligence 7:334–345. https://doi.org/10.1038/s42256-025-01017-7

Tamburrini G (2024) Artificial Intelligence and Large-Scale Threats to Humanity. In: Werthner H, Ghezzi C, Kramer J, et al. (eds) Introduction to Digital Humanism. Springer Nature Switzerland, Cham, pp 241–254. https://doi.org/10.1007/978-3-031-45304-5_16

Tessler MH, Bakker MA, Jarrett D, et al (2024) AI can help humans find common ground in democratic deliberation. Science 386:eadq2852. https://doi.org/10.1126/science.adq2852

Trammell P, Korinek A (2023) Economic Growth under Transformative AI. NBER Working Paper 31815. National Bureau of Economic Research. https://doi.org/10.3386/w31815

Turchin A, Denkenberger D (2020) Classification of global catastrophic risks connected with artificial intelligence. AI & Society 35:147–163. https://doi.org/10.1007/s00146-018-0845-5

Van Kolfschooten H (2023) The AI cycle of health inequity and digital ageism: mitigating biases through the EU regulatory framework on medical devices. Journal of Law and the Biosciences 10:lsad031. https://doi.org/10.1093/jlb/lsad031

Vargas NC (2025) Exploiting the Margin: How Capitalism Fuels AI at the Expense of Minoritized Groups. AI and Ethics 5:1871–1876. https://doi.org/10.1007/s43681-024-00502-w

Varoquaux G, Luccioni S, Whittaker M (2025) Hype, Sustainability, and the Price of the Bigger-is-Better Paradigm in AI. In: Proceedings of the 2025 ACM Conference on Fairness, Accountability, and Transparency. Association for Computing Machinery, New York, NY, USA, pp 61–75. https://doi.org/10.1145/3715275.3732006

Verdecchia R, Sallou J, Cruz L (2023) A systematic review of Green AI. Wiley Interdisciplinary Reviews: Data Mining and Knowledge Discovery 13:e1507. https://doi.org/10.1002/widm.1507

Verdegem P (2024) Dismantling AI capitalism: the commons as an alternative to the power concentration of Big Tech. AI & Society 39:727–737. https://doi.org/10.1007/s00146-022-01437-8

Wang J, Wang A, Luo K, Nie Y (2024a) Can artificial intelligence improve enterprise environmental performance: Evidence from China. Journal of Environmental Management 370:123079. https://doi.org/10.1016/j.jenvman.2024.123079

Wang P, Zhang L-Y, Tzachor A, Chen W-Q (2024b) E-waste challenges of generative artificial intelligence. Nature Computational Science 4:818–823. https://doi.org/10.1038/s43588-024-00712-6

Wang S, Cooper N, Eby M (2024c) From human-centered to social-centered artificial intelligence: Assessing ChatGPT's impact through disruptive events. Big Data & Society 11:1–14. https://doi.org/10.1177/20539517241290220

Wright D, Igel C, Samuel G, Selvan R (2025) Efficiency Is Not Enough: A Critical Perspective on Environmentally Sustainable AI. Communications of the ACM 68:62–69. https://doi.org/10.1145/3724500

Wu C-J, Raghavendra R, Gupta U, et al (2022) Sustainable AI: Environmental Implications, Challenges and Opportunities. Proceedings of Machine Learning and Systems 4:795–813.

Wynne MA, Derr L (2025) Advances in AI will boost productivity, living standards over time. Dallas Fed Economics. https://www.dallasfed.org/research/economics/2025/0624. Accessed 16 Dec 2025

Xiao T, Nerini FF, Matthews HD, et al (2025) Environmental impact and net-zero pathways for sustainable artificial intelligence servers in the USA. Nature Sustainability 8:1541–1553. https://doi.org/10.1038/s41893-025-01681-y

Xie Y, Avila S (2025) The social impact of generative LLM-based AI. Chinese Journal of Sociology 11:31–57. https://doi.org/10.1177/2057150X251315997

Zhang G, Chang F, Liu J (2024) Carbon emission prediction of 275 cities in China considering artificial intelligence effects and feature interaction: A heterogeneous deep learning modeling framework. Sustainable Cities and Society 114:105776. https://doi.org/10.1016/j.scs.2024.105776

Zhao Y, Yin D, Wang L, Yu Y (2024) The rise of artificial intelligence, the fall of human wellbeing? International Journal of Social Welfare 33:75–105. https://doi.org/10.1111/ijsw.12586

# SUPPLEMENTARY INFORMATION

## 1. Detailed methods

This section provides a detailed description of the systematic literature review method and the PRISMA flow diagram, supplementing the information provided in the Methods section of the main text. It also outlines the approach employed to estimate the approximate carbon footprint associated with the use of AI in this study.

### 1.1 Strategy

To identify the relevant literature on the impacts of artificial intelligence (AI) on environmental sustainability and human well-being, we adopted the following multi-step approach:

1. An initial search was conducted in the following databases: Scopus, arXiv, and NBER Working Papers, chosen for their comprehensive coverage of peer-reviewed articles, preprints, and economic research.
2. To complement the database search, we conducted forward and backward citation tracking to identify additional relevant studies, including screening the reference lists of included papers (backward citation tracking) and identifying subsequent works that cite them (forward citation tracking).

### 1.2 Keywords

Based on the four research questions guiding this review (see Introduction of main text), we identified two thematic groups of keywords to formulate the search strings:

1. **Environmental impacts**: *Artificial Intelligence, AI, Machine Learning, ML, Deep Learning, Large Language Models, LLMs, environment, environmental, sustainability, sustainable, carbon, biodiversity, ecological, climate change, energy, resource, impact, effect, footprint*
2. **Human well-being impacts**: *Artificial Intelligence, AI, Machine Learning, ML, Deep Learning, Large Language Models, LLMs, wellbeing, well-being, happiness, health, employment, unemployment, job, labour, labor, equality, inequality, society, impact, effect, implication*

### 1.3 Search strings

The search strings were developed based on the two keyword groups defined previously (environmental impacts; human well-being impacts). For each thematic group, we developed a specific search string. The search strings were then adapted to the syntax and functionality of each database. The process began with Scopus, which offers the most advanced search tools, and was subsequently adapted for arXiv and the NBER Working Papers (for which the advanced search was carried out through Google Scholar indexing).

The search strings were developed iteratively, with repeated refinements aimed at optimising recall and precision until satisfactory results were achieved. The final search strings for each database and thematic group are listed below.

**Scopus:**

- **Environmental impacts**: TITLE-ABS-KEY ( ( impact* OR footprint* OR effect) W/3 ( ai OR "Artificial Intelligence" OR ml OR "Machine Learning" OR "Deep Learning" OR llm* OR "Large Language Model*" ) W/3 ( environment* OR sustainab* OR ecological OR biodiversity OR carbon OR "climate change" OR energy OR resource* ) ) AND PUBYEAR > 2009 AND PUBYEAR < 2025 AND ( LIMIT-TO ( LANGUAGE , "English" ) )
  Hits: 1414 (Last access 31/12/2024)

  Notes: *The different positioning possibilities of the 3 keyword groups were considered and having both the impact-related keyword group and the environment-related keyword group close to the AI-related keyword group returned more results, encompassing the largest number of phrases relevant to the research question.*

- **Human well-being impacts:** TITLE-ABS-KEY ( ( impact* OR implication* OR effect) W/3 (ai OR "Artificial Intelligence" OR ml OR "Machine Learning" OR "Deep Learning" OR llm* OR "Large Language Model*") W/3 ( wellbeing OR well-being OR happiness OR health OR *employment OR labor OR labour OR job* OR *equality OR society) ) AND PUBYEAR > 2009 AND PUBYEAR < 2025 AND ( LIMIT-TO ( LANGUAGE,"English"))
  Hits: 1449 (Last access 31/12/2024)

  Notes: *Alternative terms referring to AI's social and well-being impacts, such as life satisfaction, time use, work, equity, community, and social relations, were tested. The most complete combination of words yielding relevant results was finally selected.*

  Since the other databases do not support proximity operators, the operator W/ was replaced by AND. To restrict the search to articles discussing impacts of artificial intelligence, the AI-related keyword group was searched in the title.

**arXiv**

- **Environmental impacts**: date_range: from 2010-01-01 to 2024-12-31; AND title=ai OR "Artificial Intelligence" OR ml OR "Machine Learning" OR "Deep Learning" OR llm OR "Large Language Model"; AND abstract=environment OR environmental OR sustainable OR sustainability OR ecological OR biodiversity OR carbon OR "climate change" OR energy OR resources; AND abstract=impact OR footprint
  Hits: 1,412 (Last access 31/12/2024)
  1,342 retrieved through arXiv API

  Notes: *The word "effect" was excluded because the arXiv search engine does not strictly enforce exact phrase matching, resulting in unwanted matches such as "effectiveness", "effectively", and "effective", and consequently retrieving many irrelevant records.*

- **Human well-being impacts:** date_range: from 2010-01-01 to 2024-12-31; AND title=ai OR "Artificial Intelligence" OR ml OR "Machine Learning" OR "Deep Learning" OR llm OR "Large Language Model"; AND abstract=wellbeing OR well-being OR happiness OR health OR employment OR unemployment OR labor OR labour OR job OR equality OR inequality OR society; AND abstract=impact OR implication
  Hits: 1,072 (Last access 31/12/2024)
  927 retrieved through arXiv API

**NBER working papers (via Google Scholar indexing)**

- **Environmental impacts**: site:nber.org (intitle:(ai OR "Artificial Intelligence" OR ml OR "Machine Learning" OR "Deep Learning" OR llm OR "Large Language Model" )) AND( impact OR footprint OR effect ) AND ( environment OR environmental OR sustainable OR sustainability OR ecological OR biodiversity OR carbon OR "climate change" OR energy OR resources )
  Date: 2010-2024
  Hits: 948 (Last access 31/12/2024)
  652 retrieved through Zotero

- **Human well-being impacts:** site:nber.org (intitle:(ai OR "Artificial Intelligence" OR ml OR "Machine Learning" OR "Deep Learning" OR llm OR "Large Language Model")) AND ( impact OR effect OR implications) AND ( wellbeing OR well-being OR happiness OR health OR employment OR unemployment OR labor OR labour OR job OR equality OR inequality OR society )
  Date: 2010-2024
  Hits: 930 (Last access 31/12/2024)
  646 retrieved through Zotero

  Notes: *The output records (948 and 930 respectively) included many papers published before 2010. The number of retrieved studies reported is net of deleted records falling outside the 2010-2024 period.*

## 1.4 Eligibility criteria

To determine the eligibility of the studies for inclusion, the following criteria were applied both during the initial selection and during the analysis of the full-text studies:

- Publication date: published in 2010 or later.
- Type of source: peer-reviewed articles, conference proceedings, working papers, preprints, or grey literature from reputable institutions (e.g., reports from major universities, European Institutions, or United Nations agencies).
- Language: available in English.
- Accessibility: full text must be available.
- Relevance: addresses one or more of the four research questions of the systematic review (see Introduction).

## 1.5 Data management

The full electronic search strategy for each major database, including any applied limits, is documented in the systematic review protocol to ensure reproducibility. All information sources and the date of the last search are recorded in the protocol and summarised in the PRISMA flow diagram (Supplementary Fig. 1). The PRISMA flow diagram also reports the exact number of records retrieved, screened, included, and excluded at each stage of the review process.

Bibliographic references were managed using Zotero. Screening and data extraction were carried out using structured Excel spreadsheets. All project files were stored in a shared folder on Teams to ensure secure access and version control within the project team. The systematic review protocol is publicly available on protocols.io to ensure transparency (Carmeno et al. 2026).

## 1.6 Selection process

For each record retrieved, metadata (i.e. author(s), title, year, journal, link, and abstract) were exported to a master spreadsheet and used for the title-and-abstract screening stage. Prior to screening, duplicate records were identified and removed (matching on title/DOI and cross-checking author–year combinations). Screening was conducted manually. To reduce potential selection bias, screening was conducted blind to authorship and publication source during both the title-and-abstract and full-text stages.

After the initial screen, full texts of potentially relevant records were obtained and assessed against the predefined eligibility criteria. At this stage, explicit reasons for exclusion were recorded for all studies not meeting the criteria. Included studies were classified by relevance to our research questions into three categories: *very relevant* (the study's core contribution directly addresses one or more research questions), *relevant* (the study addresses one or more research questions, but this is not its core contribution), and *less relevant* (the study tangentially addresses or intersects with one or more research questions).

Throughout the process, a screening spreadsheet was maintained to log decisions at each step, ensuring auditability and transparency. Supplementary Fig. 1 reports the PRISMA flow diagram depicting the number of records identified, deduplicated, screened, assessed for eligibility, included, and excluded (with reasons) across all stages of the review.

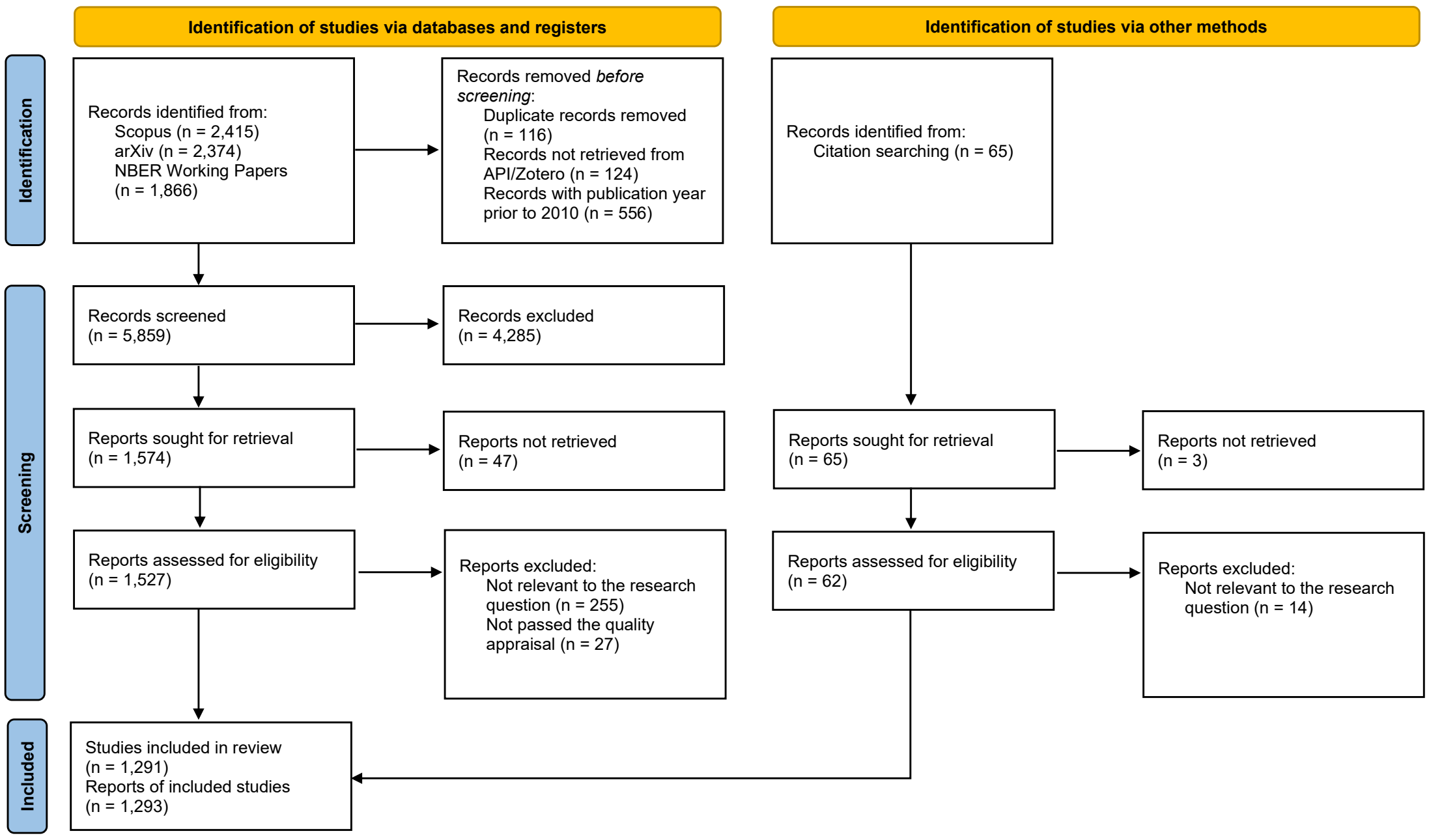

**Supplementary Fig. 1 PRISMA 2020 flow diagram.** The figure illustrates the identification, screening, and selection of studies in our systematic literature review

## 1.7 Data extraction

Data extraction was carried out using a structured spreadsheet, developed iteratively during an initial pilot phase. The extraction form combined closed categorical fields with free-text fields for summary, key findings, methods, limitations, research gaps and notes. This hybrid structure ensured consistency in coding for quantitative analysis, whilst allowing the flexibility needed to capture context-specific elements of the study for qualitative analysis.

The first set of fields included bibliographic and descriptive information: author(s), title, year of publication, journal or source, link/DOI, and a brief summary of the study's objectives, methods, and key findings.

These fields were followed by standardised categorical variables used to classify each study for the quantitative analysis:

- Research line: AI environmental impact, AI well-being impact. Studies could be cross-coded if they addressed both.
- Scale of analysis: macro (global, cross-country); macro (national, sub-national); micro (enterprise; set of projects/models); micro (application/model/hardware).
- Broad method: conceptual/theoretical, qualitative, or quantitative. If a study used multiple methods, it was categorised according to the predominant method or the method that had the greatest influence on the results.
- Specific method:
    - Conceptual/theoretical: Conceptual Framework, Scoping/Literature Review, Conceptual System Design, Ethical Analysis, Policy Analysis, Perspective Analysis.
    - Qualitative: Qualitative Experimental Evaluation, Interview, Survey, Case Study, Participatory Method, Thematic/ Content Analysis, Expert Elicitation.
    - Quantitative: Quantitative Experimental Evaluation, Quantitative Survey, Quantitative Case Study, Statistical Analysis, Macroeconomic Analysis, Life Cycle Assessment, Optimisation, Remote Sensing.
- Impact type:
    - Environmental: energy use, material use, water use, land use, ecological footprint, $CO_2$ emissions, other pollutants, biodiversity, Sustainable Development Goals (SDGs).

- Well-being: income, employment, inequality, health, subjective well-being, cognitive capabilities, social cohesion, existential risk.

  Studies could be cross-coded if they addressed multiple impacts.
- Level at which impacts arise, following Kaack et al.'s (2022) framework categories: computing-related, application-level, or systemic. For environmental studies at the application and systemic levels, we also coded the most recurring impact pathways:
    - Application-level: improvements in firms' environmental performance, or other.
    - Systemic: rebound effects, shifts in consumer behaviour, or other.
- Sentiment: positive, negative, or neutral. The sentiment was assigned at the study level based on the authors' own assessment of the impact. If a study reported only benefits and no harm, it was classified as positive. Conversely, if it reported only harm and no benefits, it was classified as negative. If both aspects were reported without a clear overall trend, or if the description was explicitly ambivalent, the study was classified as neutral. Where both benefits and harms were reported, but the authors indicated an overall trend, for example in the abstract or discussion, we followed that indication.

The different categories were developed through an in-depth review of the literature, aimed at capturing the full range of approaches and impacts used in the various studies. This was a highly iterative process, during which the classification scheme was systematically updated and expanded whenever new methodological characteristics or impact dimensions were identified, thereby ensuring a robust, comprehensive, and internally consistent taxonomy.

Finally, the data extraction form included the following open-text fields, designed to facilitate an in-depth qualitative analysis:

- Key findings: a summary of the main findings, conclusions or implications of the study.
- Methods: a detailed description of the study's assumptions, data sources, analytical approach, and the specific method used.
- Limitations: a summary of the main limitations associated with the study's design or method, either as reported by the authors or identified during full-text screening and quality appraisal.
- Research gaps: any gaps in the currently available evidence or areas requiring further investigation, as explicitly highlighted by the study's authors.
- Notes: additional comments intended to supplement the coded fields, clarify specific cases, or provide relevant insights for the final summary.

For studies classified as very relevant and relevant, data extraction was performed manually. For studies classified as less relevant (approximately 30% of the total), data extraction was supported by a Large Language Model (LLM), specifically Microsoft 365 Copilot Enterprise (GPT-4 Turbo). LLM-assisted data extraction was performed using two standardised prompts. Each prompt was developed iteratively, monitoring the quality of the output until the expected result was achieved. Before including the extracted information in the data extraction spreadsheet, each individual datum was verified against the content of the original study.

### 1.7.1 Operational definitions of impact types

To ensure consistent coding, we have applied the following definitions when coding the field relating to the impact type in the environmental and well-being studies.

**Environmental impacts**

- Energy use: electricity used by AI development, training, inference, and supporting infrastructure, or by any related process across the supply chain (e.g. hardware manufacturing, trade).
- $CO_2$ emissions: greenhouse-gas emissions ($CO_2$ or $CO_2$-equivalent) attributable to AI activities or infrastructure.
- Water use: freshwater withdrawals or consumption across onsite cooling and upstream electricity and supply-chain stages.
- Material use: critical minerals and other material inputs for AI hardware and infrastructure across supply chains.

- Land use: land occupation or change linked to data centres, infrastructure siting, mining, or related supply-chain activities.
- Biodiversity: impacts on species, habitats, or ecosystem integrity associated with AI infrastructure and supply chains.
- Other pollutants: non-$CO_2$ greenhouse gases (GHGs) and air, water, or solid-waste pollutants, including toxics and e-waste.
- Ecological footprint: biologically productive land and water area required to provide the resources consumed by a population and to assimilate its wastes ($CO_2$), summed over cropland, grazing land, fishing grounds, forest products land, built-up land, and carbon demand on land, and reported in global hectares following the National Footprint & Biocapacity Accounts method (Lin et al. 2018).
- Sustainable Development Goals (SDGs): Environmentally-related SDG indicators used to assess AI impacts.

**Well-being impacts**

- Employment: effects of AI on jobs and tasks, including displacement or creation, augmentation and productivity, job quality (autonomy, workload, meaning), and the use of algorithmic management or microwork platforms.
- Income: changes in wages and earnings attributable to AI adoption, including wage dispersion or compression, returns to top talent or capital, and income volatility in contingent or on-demand work.
- Inequality: distributional impacts across groups, sectors or regions linked to AI, including access and literacy divides, bias in allocation and representation, concentration of economic gains, data-ownership asymmetries, and North–South gaps.
- Health: clinical or public-health outcomes and access to care influenced by AI (diagnosis, treatment planning, triage, mental-health support), including differential benefits and harms and risks related to fairness, privacy, consent, and liability.
- Subjective well-being: effects on life satisfaction, happiness, and affect associated with AI use, including techno-stress or empowerment, over-dependence and detachment, social connection versus isolation, time-use shifts, and perceived autonomy.
- Cognitive capabilities: impacts on attention, memory, critical thinking, problem-solving, and creativity, including cognitive offloading, reduced deep engagement, homogenisation of language and thought, and longer-term “cognitive debt”.
- Social cohesion: changes in trust, authenticity norms, and community bonds associated with AI-mediated content and systems, including manipulation and misinformation, surveillance and social scoring, marginalisation of minority voices, and power concentration.
- Existential risk: high-impact risks linked to advanced AI which could threaten the survival of human beings, including misalignment and loss of control, misuse or jailbreaking for bio or cyber capabilities, undermining of nuclear stability, and autonomous weapons or AI-enabled warfare.

SDG-based well-being studies were mapped to the specific well-being impact type they operationalise (e.g. SDG 3 maps to health impacts, SDG 10 maps to inequality impacts). To avoid a generic category and double counting, no separate SDG category is used in the well-being domain. By contrast, in the environmental domain, SDG indicators often aggregate across several environmental pressures and do not map cleanly to our environmental impact types. Accordingly, to preserve comparability and avoid inflating single categories, we agreed on including a separate “SDGs” category for environmental studies. Where a study reports both SDG indicators and a specific environmental metric (e.g. $CO_2$), it is coded to both.

### 1.7.2 Operational definitions of levels at which impacts arise

Following Kaack et al.'s (2022) framework for greenhouse gas emissions, and expanding it to cover all the environmental impact types included in our taxonomy (see Section 1.7.1), the following definitions have been applied to consistently categorise environmental studies by the level at which impacts arise:

- Computing-related studies address model, hardware, and infrastructure impacts, including training, inference, data-centre operations, mining, hardware manufacturing, trade, and end-of-life disposal.
- Application-level studies address impacts arising from AI use in specific sectors or use-cases.
- Systemic studies address indirect, longer-term impacts such as rebound effects, behavioural or institutional shifts.

Following our expansion of the framework (see Fig. 5 of the main text), the following definitions were applied in categorising well-being studies by the level at which impacts arise:

- Computing-related studies address impacts arising in the upstream supply chain of AI, including data consent gaps, data workers' labour conditions, ecologically unequal exchange linked to hardware manufacturing, and impacts on local communities from infrastructure siting.
- Application-level studies analyse the impacts resulting from the implementation of AI in specific industries and related to the practical functions performed by this technology.
- Systemic studies address indirect, longer-term, economy-wide, societal or cultural effects such as concentration of wealth and power, institutional legitimacy loss, acceleration of scientific discovery, and enhanced crisis response.

Additionally, for environmental studies classified as application-level or systemic, we coded the predominant impact pathway emphasised in the study, and the following definitions applied:

- Improvements in firms' environmental performance: studies addressing environmental impacts primarily in terms of changes in firms' operational performance, such as energy efficiency, emissions intensity, or resource use.
- Rebound effects: studies addressing indirect environmental impacts whereby efficiency gains, cost reductions, or expanded capabilities enabled by AI lead to increased activity, consumption, or systemic responses that may offset expected environmental benefits.
- Shifts in consumer behaviour: studies addressing environmental impacts arising from AI-mediated changes in consumption patterns, demand, or user behaviour, such as AI-driven targeted advertising increasing consumption.

### 1.7.3 Operational definitions of method categories

We have applied the following definitions to ensure consistent coding of the methods applied in the surveyed studies.

**Conceptual/theoretical**

- Conceptual framework: development or refinement of conceptual or theoretical constructs, mechanisms, or relationships without new empirical analysis.
- Scoping/literature review: systematic mapping or synthesis of prior work using explicit search, inclusion, and selection procedures.
- Conceptual system design: specification of a software architecture, system design, or processing pipeline, including components and data or process flows, without empirical benchmarking.
- Ethical analysis: normative examination of AI impacts, risks, or trade-offs through the application of ethical theories, principles, or frameworks.
- Policy analysis: assessment of policy options, instruments, or governance mechanisms in relation to a specified objective or societal challenge.
- Perspective analysis: viewpoint or commentary advancing a thesis or agenda without systematic evidence gathering.

**Qualitative**

- Qualitative experimental evaluation: lab or field evaluation using qualitative outcomes or protocols where inferential statistics are not the primary analysis.
- Interview: structured, semi-structured, or open interviews analysed qualitatively.

- Survey (qualitative): open-ended or mixed-item surveys where qualitative analysis is the primary method.
- Case study: in-depth examination of a bounded case (e.g. firm, project, community) using qualitative evidence.
- Participatory method: research method that views project participants or stakeholders as active agents, including through co-design, workshops, or citizens' assemblies.
- Thematic/content analysis: the systematic coding of texts, transcripts, or multimedia materials in order to identify themes or categories.
- Expert elicitation: structured elicitation of expert judgements, or a Delphi process, with qualitative synthesis.

**Quantitative**

- Quantitative experimental evaluation: controlled experiments or direct measurements quantifying AI's environmental impacts (e.g. energy use or emissions during training, inference, or operation) using instruments such as power meters or carbon trackers.
- Quantitative survey: survey with primarily quantitative items analysed statistically.
- Quantitative case study: single- or few-case design where quantitative indicators and statistical analysis are the primary method.
- Statistical analysis: descriptive or inferential analysis of observational data, including econometric techniques (e.g. regression, panel, time-series, spatial models, matching, synthetic controls). Excludes economy-wide macroeconomic models, LCA, optimisation studies, and direct metering-based measurement.
- Macroeconomic analysis: economy-wide models (e.g. input–output / MRIO / EEIO, CGE, DSGE) to evaluate AI's impacts at a regional, national, or global scale.
- Life Cycle Assessment (LCA): application of LCA method to quantify AI systems' impacts across the life cycle (hardware, deployment, operation, end-of-life) with defined functional unit, boundaries, inventory, and impact method.
- Optimisation: development, evaluation, or application of methods that improve the energy, resource, or compute efficiency of AI systems (e.g. compression, efficient architectures, power capping) to reduce energy use or $CO_2$ emissions.
- Remote sensing: analysis of satellite, aerial, or other remotely sensed data to infer environmental impacts associated with AI infrastructure or AI-enabled applications (e.g. land or resource use).

## 1.8 Quality assessment

At the review level, to ensure internal validity and reproducibility of the review, we followed the PRISMA 2020 framework (Page et al. 2021). We followed the PRISMA checklist and guidelines carefully and structured the systematic review protocol accordingly. To ensure transparency, the systematic review protocol is publicly available on protocols.io (Carmeno et al. 2026).

At the level of the individual studies included, quality assessment was conducted during the full-text screening stage. Screening, quality appraisal, and data extraction were conducted using a predefined codebook and decision rules. Each study was examined in detail to evaluate the soundness of its methods, including the clarity of its assumptions, the suitability of its analytical approach, and the coherence between its data, methods, and conclusions. During this assessment, we also examined potential sources of bias, such as insufficient data support, unclear causal links or causal language without an identification strategy, undisclosed assumptions that materially affect results, unvalidated proxies or measurements, circular reasoning between inputs and outcomes, reuse of overlapping datasets without disclosure, and selective reporting of results. Studies judged to be at critical risk of bias were excluded from the review and these exclusions are reported in the PRISMA flow diagram together with the reasons for exclusion.

To ensure a structured, reproducible, and comprehensive data extraction, our procedures were based on iterative cycles of literature review, definition of analysis categories, analysis of the literature in relation to the defined categories, and revision of the categories. By doing so, we tested various categorisations and coding methods until the most appropriate ones were identified.

For studies classified as less relevant, LLM-assisted extraction of textual outputs and categorisations were verified datum by datum against the original PDFs before entry. The quality of the outputs was thoroughly checked during the prompt definition phase, which was an iterative process in which the prompt was refined at each stage until the desired quality and level of detail were achieved. Once finalised, a fixed prompt set was used for extraction. For data extracted with LLM assistance, the quality control checklist included an item for each piece of extracted information, verifying its accuracy and cross-checking it against the original text. LLM-assisted extraction was conducted within our Microsoft 365 Enterprise environment and prompts and outputs were versioned and stored within this environment. No extracted information was accepted without manual verification.

As an internal consistency check, we carried out spot checks on category tags on a random subset of records, cross-checking data with the source PDFs. This check found no discrepancies.

## 1.9 Synthesis

The final data extraction spreadsheet served as the basis for the synthesis phase. The quantitative synthesis was based on the closed-ended categorical fields in the extraction form. These variables were analysed using descriptive statistics to identify and visualise patterns relating to impact type, the level at which impacts arise, analytical scale, methods, and sentiment. The results obtained were examined for relevance, clarity, novelty, and interpretative value. Only the most informative and robust results, particularly those offering new insights or significant empirical patterns, were selected for inclusion in the final manuscript. Additional findings, which provide further nuance, are reported in the Supplementary Information.

The qualitative synthesis was based on the free-text fields compiled during the data extraction phase. The key findings extracted were examined to identify the most recurrent items, emerging themes, as well as points of convergence and divergence among the studies. The most recurrent findings within each level and type of impact, characterised by the higher estimated impact severity and the higher degree of consensus and certainty, were selected for inclusion in the main manuscript. The complete list of recurring findings is presented in Section 3.1, in Supplementary Table 1 for environmental impacts and in Supplementary Table 2 for impacts on well-being.

Data on research methods, limitations, and gaps were analysed to produce a structured overview of the strengths and weaknesses of the current evidence base. These qualitative considerations helped to establish an overall assessment of research gaps and future priorities, complementing and expanding on the patterns identified in the quantitative synthesis.

## 1.10 $CO_2$ emissions statement

To roughly estimate the $CO_2$ emissions implied by this study's use of a large language model (LLM), we estimated the tokens and associated energy use required for two standardised prompts and replies on three studies per iteration for 130 iterations. As official data on energy use of Microsoft Copilot are not publicly available, we used data available in the literature and relied on a number of assumptions.

Below are the main assumptions and calculations:

Assuming 1 word = 1.33 tokens (OpenAI 2025)

- Prompt 1 = 831 words × 1.33 ≈ 1,105 tokens
- Reply 1 = 500 words on average × 1.33 ≈ 665 tokens
- Attachments = 3 PDFs × 4,500 words on average × 1.33 ≈ 17,955 tokens
- Prompt 2 = 38 words × 1.33 ≈ 51 tokens
- Reply 2 = 160 words on average × 1.33 tokens ≈ 213 tokens
- Total tokens of one iteration = 1,105 + 665 + 17,955 + 51 + 213 = 19,989 tokens ≈ 20,000 tokens

Assuming average energy use per token = 0.05 Wh/token[i]

- Energy use of one iteration = 20,000 tokens × 0.05 Wh/token = 1,000 Wh
- Total energy use = 1,000 Wh × 130 iterations = 130,000 Wh = 130 kWh

Assuming an average European cloud-mix electricity emission factor = 0.3 kg $CO_2$/kWh (Bastos et al. 2024)

- $CO_2$ emissions = 130 kWh × 0.3 kg $CO_2$/kWh = 39 kg $CO_2$

All estimates are indicative and sensitive to assumptions regarding tokenisation, model efficiency, and electricity mix.

## 2. Quantitative results

This section contains notes on the figures included in the main text, intended to supplement the information provided and ensure that the figures can be fully reproduced.

**Fig. 1 Distribution of studies by impact type and level at which impacts arise**

Panel a shows environmental studies organised by environmental impact type and panel b shows well-being studies organised by well-being impact type. The complete list of impact types and their definitions is provided in Section 1.7.1.

Horizontal bars in panels a and b show totals of study–impact pairs over the study period. Counts are absolute numbers of study–impact pairs. A study can be counted in multiple impact types if it addresses more than one type. Multiple versions of the same study were collapsed to the most complete version before counting. If a study covers both domains (i.e. environmental and well-being impacts) it appears in each relevant panel. Panels a and b share the same x-axis range to enable visual comparability. Categories with very few items should be interpreted cautiously (e.g. land use).

Panels c and d show the percentage of studies by the level at which impacts arise. Pie chart wedges show percentages of unique studies at each level within the relevant corpus. The complete list of levels at which impacts arise and their definitions is provided in Section 1.7.2. Multiple versions of the same study were collapsed to the most complete version before counting. Percentages sum to 100% within each panel and are rounded to whole numbers. Very small slices should be interpreted cautiously (e.g. computing-related well-being studies).

**Fig. 2 Percentage of studies by broad method used and analytical scale**

Broad method categorisation follows the taxonomy reported in Section 1.7.3 (conceptual/theoretical, qualitative, quantitative). The scale of analysis was first assigned to four categories: macro (global, cross-country), macro (national, sub-national), micro (enterprise, set of projects/models), and micro (application/model/hardware). These four categories were then merged into macro and micro to develop the final figure. Across panels, percentages sum to 100% per bar and are rounded to whole numbers. Multiple versions of the same study are collapsed to the most complete version before counting.

**Fig. 3 Number of studies by specific method used and impact type**

In the two heatmaps shown in this figure, each cell shows the number of studies using a given specific method to assess a given impact type across the corpus. Panel a shows the heatmap relating to studies on environmental impacts, while panel b shows the heatmap relating to impacts on human well-being. The vertical axis of both panels shows the specific method used (defined in

[i] Estimates of per-token energy use reported in the literature span several orders of magnitude (≈ 0.000003–0.069 Wh/token), depending on model size, hardware, batching, and utilisation, and are based on open-source models (Niu et al. 2025). While proprietary models may achieve higher per-token efficiency through specialised hardware and optimisation, they are also substantially larger and embedded in more complex inference pipelines, making net energy use per token uncertain. In the absence of provider-specific data, a conservative high-end point estimate is therefore used to avoid underestimating emissions.

Section 1.7.3), while the horizontal axis shows the impact type considered (defined in Section 1.7.1). A study can be counted in multiple impact types. Shading encodes absolute counts within each panel. Legends are panel-specific. No normalisation is applied across panels. Methods and impact types are ordered by total counts within each panel for readability. Multiple versions of the same study are collapsed to the most complete version. Cells with very low counts should be interpreted cautiously.

**Fig. 4 Percentage of studies by overall sentiment**

Sentiment is assigned at the study level from the authors' own reporting. A study is coded positive when it reports benefits and no harms, negative when it reports harms and no benefits, and neutral when both are presented without a clear overall tendency. Where both directions are reported but the authors state an overall direction in the abstract or discussion, we follow that stated direction.

Multiple versions of the same study are collapsed to the most complete version. Percentages are rounded to whole numbers and sum to 100% within each bar. Categories with very few items should be interpreted cautiously.

# 3. Qualitative results

This section presents the complete lists of findings and research gaps that constitute the main outcome of the qualitative analysis in our systematic review. It also includes some brief methodological notes on the selection of studies for the creation of the summary tables included in the main text (i.e. Tables 1 and 2).

## 3.1 Extended tables of findings and research gaps

Supplementary Table 1 provides the complete list of recurring findings and research gaps on the environmental impacts of AI, with references. The list of findings was compiled through data extraction, clustering by the level at which impacts arise, identification of recurring findings, and derivation of ranges or means where appropriate, depending on indicator type and available evidence. The list of research gaps combines gaps reported in the reviewed literature — which were extracted, clustered, and curated — with those identified through the quantitative analysis.

**Supplementary Table 1 Full findings and research gaps on AI's environmental impacts.** The table reports the complete list of recurring findings and underexplored topics organised by the level at which impacts arise. Table 1 of the main text provides a summarised version

| Level | Full findings | Research gaps |
|---|---|---|
| ***Computing-related impacts*** | • Escalating computing requirements drive increasing energy use and $CO_2$ emissions for diminishing accuracy gains: ~4–6% accuracy gain from GPT-2 to GPT-4 with ~2,000× more compute (Varoquaux et al. 2025).<br>• Inference emissions exceed training emissions at deployment scale (de Vries 2023; Natarajan et al. 2022), with GPT-3 providing an estimate of ~25× higher inference emissions (Chien et al. 2023).<br>• As of March 2024, ChatGPT was estimated to consume more than half a million kilowatt-hours of electricity each day, nearly equal to the consumption of 180,000 U.S. households (Kshetri 2024).<br>• Efficiency techniques can reduce $CO_2$ emissions by up to ~90%, but the electricity grid mix remains the dominant factor (Alswaitti et al. 2024; Dodge et al. 2022; Gibney 2022; Jagannadharao et al. 2023; Luccioni and Hernandez-Garcia 2023; Verdecchia et al. 2022).<br>• Embodied emissions from hardware and infrastructure may account for ~30–50% of life-cycle carbon footprint (Jiang et al. 2024; Kaack et al. 2022; Wright et al. 2025; Wu et al. 2022).<br>• Greenhouse gas emissions from major AI firms rose in recent years driven by data centre expansion that outpaces climate pledges e.g. Google's 2023 emissions increased 13% year-over-year, while Microsoft's emissions have increased 29% since 2020 (Guidi et al. 2024).<br>• AI workloads consume ~10–50 TWh/year and may reach 200–400 TWh/year by 2030, then comprising ~35–50% of global data centre electricity use (Kamiya and Coroamă 2025).<br>• Data centres are increasingly located in regions with high carbon intensity; for instance, ~95% of new US facilities are in areas with above-average carbon intensity (Guidi et al. 2024). | • Most studies focus on training; inference, infrastructure and hardware manufacturing, and disposal phases are often neglected.<br>• Most studies focus on energy and $CO_2$ emissions with limited attention to water, material, other pollutants, and biodiversity impacts.<br>• Regional and community-level impacts are poorly understood.<br>• Firm-level environmental disclosures are limited, non-standardised, and rarely extend beyond energy use and carbon emissions.<br>• Environmental sustainability metrics are rarely integrated into benchmarking, model evaluation, or reporting standards. |

| Level | Full findings | Research gaps |
|---|---|---|
| | • Data centres are built in water-stressed regions, competing with local communities for fresh water and electricity and generating environmental conflicts (Lehuedé 2025).<br>• AI's water usage encompasses on-site water use for server cooling under scope 1, off-site water use for electricity generation under scope 2, and supply-chain water use for server manufacturing under scope 3 (Li et al. 2025).<br>• Training a single LLM can use over 5 million litres of freshwater, while inference can require ~500 ml for every 10–50 responses (Li et al. 2025).<br>• Carbon optimisation strategies may further increase water use (Li et al. 2025).<br>• Global AI demand is projected to withdraw 4.2–6.6 billion $m^3$ of water by 2027 from scope 1 and 2 alone (Li et al. 2025).<br>• Semiconductor fabrication uses millions of litres of ultrapure water daily and generates hazardous waste (Frost and Hua 2019). Hardware manufacturing drives human toxicity, mineral and metal resource use, and freshwater eutrophication impacts, accounting for about 80% of these impacts on average for GPT-4 training using Nvidia A100 GPUs (Falk et al. 2025).<br>• Mining for AI-critical materials such as lithium, cobalt, and rare earths causes water depletion, pollution, and biodiversity loss (Lehuedé 2025; Wright et al. 2025).<br>• AI accelerates hardware turnover, contributing to global e-waste (Wang et al. 2024b), which is projected to reach ~75 Mt annually by 2030 (Forti et al. 2020), and exposing soil and water to toxic components such as lead, mercury, and cadmium.<br>• LLM server hardware alone could add ~0.8–2.5 Mt $yr^{-1}$ of e-waste by 2030 (cumulative 1.2–5.0 Mt over 2020–2030), subject to deployment and refresh-rate assumptions (Wang et al. 2024b).<br>• As of May 2025, 84% of the use of leading LLMs occurred on models with no environmental disclosure (Luccioni et al. 2025a). | |
| ***Application-level impacts*** | • AI can improve firms' energy and resource efficiency, with an additional industrial robot per 100 workers reducing energy intensity by ~2.5% (Zhang and Zeng 2024).<br>• Environmental benefits vary by sector, region, firm-type, and deployment stage, with stronger effects in high-income, industry-dense and highly regulated areas (Chen et al. 2024; Fu et al. 2024; Wang et al. 2024a; Zhang et al. 2024), and in sectors like energy, manufacturing, agriculture, smart cities, and environmental monitoring (Ghobakhloo et al. 2024; Li et al. 2024; Liu et al. 2022; Zürner et al. 2023).<br>• AI may increase emissions by reinforcing carbon-intensive sectors such as fossil fuels and heavy industry through improved extraction and logistics, and by stimulating AI-driven personalised consumption through targeted advertising and recommender systems (Gaffney et al. 2025).<br>• AI may facilitate energy transitions, though observed effects remain small, with 1% AI growth linked to a 0.0025% increase in the renewable-energy share of total final energy consumption (Wang et al. 2024c). | • AI's role in carbon-intensive sectors is understudied.<br>• Green and brown AI impacts are rarely compared.<br>• Studies rely heavily on proxies (AI is measured as the number of industrial robots) and linear assumptions.<br>• Causal pathways remain poorly understood.<br>• Research is concentrated in China and high-income countries. |
| ***Systemic impacts*** | • Efficiency gains in AI are triggering the rebound effect: efficiency improves yet aggregate electricity use continues to grow (de Vries 2023; Kaack et al. 2022; Luccioni et al. 2025b; Willenbacher 2021; Wright et al. 2025).<br>• Efficiency savings may be spent on resource-intensive upgrades, such as smartphones with new AI features (de Vries 2023; Luccioni et al. 2025b).<br>• AI may enable broader economic growth, including through power purchase agreements, increasing the overall burden on the environment (Luccioni et al. 2025b; Willenbacher 2021).<br>• Substitution of physical products and processes (e.g. AI-generated art), scaling to larger models, and shifting workloads to hyperscale data centres may raise infrastructure and material demands (Luccioni et al. 2025b).<br>• AI-driven efficiency gains in other sectors, such as transport, logistics, and manufacturing, may induce sectoral rebound effects (Luccioni et al. 2025b).<br>• AI-driven personalisation and advertising may intensify consumption (de Vries 2023; Kaack et al. 2022; Luccioni et al. 2025b; Willenbacher 2021).<br>• AI might enhance long-term climate planning and decision-making, supporting climate-change mitigation.<br>• AI might lock in existing unsustainable socio-technical configurations and reinforce path dependency e.g. private car-centric transport through autonomous vehicles (de Vries 2023; Gossen and Lell 2023; Kaack et al. 2022; Luccioni et al. 2025b; Willenbacher 2021).<br>• Market incentives might prioritise scalable, high-profit applications, such as advertising and fossil fuel optimisation, over sustainability-oriented ones (Rikap 2024; Varoquaux et al. 2025).<br>• AI development is increasingly aligned with corporate interests, which risks reinforcing extractive models and delaying structural reforms (Rikap 2024; Varoquaux et al. 2025). | • No empirical studies currently quantify systemic environmental impacts of AI; existing literature remains largely conceptual.<br>• Economic rebound effects from AI efficiency gains remain unquantified, with no robust estimates of current or projected impacts.<br>• AI-driven shifts in culture and behaviour, lock-in dynamics, and path dependency are underexplored.<br>• The literature lacks methods for assessing system-level environmental impacts, pointing to a need for tailored analytical approaches. |

Following the same method, we developed the lists of findings and research gaps concerning the impacts of AI on human well-being. Unlike with environmental impacts, we grouped and selected the recurring impacts and research gaps based on the types of impact considered. Supplementary Table 2 below presents the resulting lists by impact type, with references.

**Supplementary Table 2 Full findings and research gaps on AI's impacts on human well-being.** The table reports the complete list of recurring findings and underexplored topics organised by impact type. Table 2 of the main text provides a summarised version

<table>
<tr><th>Impact type</th><th>Full findings</th><th>Research gaps</th></tr>
<tr><td>Social cohesion</td><td>• AI may reshape social relationships by fostering parasocial bonds and emotional dependence, potentially displacing real-world relationships and weakening community cohesion (Li and Pang 2024).<br>• AI-mediated communication, synthetic content, and automated decision systems may alter norms of authenticity, responsibility, and agency, challenging shared moral expectations and social accountability (Li and Pang 2024; Mitra et al. 2025).<br>• AI can support consensus-building in large-scale decisions, helping to facilitate participatory processes and citizens' assemblies (McKinney 2024; Tessler et al. 2024).<br>• Algorithmic amplification, micro-targeted political messaging, and synthetic content may distort public discourse, fuel misinformation and polarisation, and erode social trust in institutions (Mitra et al. 2025; Pappalardo et al. 2024).<br>• Synthetic media and the increasing liar's dividend may undermine evidence authentication and public understanding, threatening judicial and electoral integrity and the credibility of the information ecosystem (Mitra et al. 2025; Shoaib et al. 2023).<br>• Synthetic media can be used to create non-consensual or misleading depictions of individuals, facilitating harassment, gender-based digital violence, and reputational harm (Birhane 2022; Shoaib et al. 2023).<br>• AI models optimise for patterns and averages, risking the exclusion of minority perspectives and dissenting positions, thereby weakening democratic deliberation (Bender et al. 2021; Mohamed et al. 2020).<br>• Cultural and linguistic biases in AI systems may marginalise non-Western perspectives, compromising epistemic diversity and reinforcing global hierarchies (Bender et al. 2021; Mohamed et al. 2020).<br>• AI-enabled surveillance, facial recognition, and predictive policing may intensify social control and infringe on privacy, autonomy, and civic freedoms (Hussein et al. 2021; Kalluri et al. 2023; Vargas 2025).<br>• Rising computational and financial barriers risk excluding smaller players from frontier AI research, undermining pluralism (Rikap 2024; Varoquaux et al. 2025).<br>• AI development dominated by a few Big Tech firms risks concentrating techno-economic power, reinforcing oligopolistic control over data, infrastructure, and dual-use technologies and jeopardising democratic oversight (Benjamins and Salazar 2020; Coveri et al. 2025; Rikap 2024; Verdegem 2024).<br>• A global AI arms race may strengthen the digital-military complex and consolidate geopolitical power among a few state-corporate alliances, increasing conflict risks and societal insecurity (Coveri et al. 2025).</td><td>• Limited empirical evidence on social cohesion impacts.<br>• Lack of research on long-term impacts on social relationships.<br>• Limited evidence on how AI affects public discourse, consensus-building, and democratic processes.<br>• Limited analysis of surveillance and repression risks in AI deployment across political regimes.<br>• Few empirical studies on concentration of power and the digital-military complex.</td></tr>
<tr><td>Inequality</td><td>• AI may shift income from labour towards capital and frontier firms, concentrating gains among a small number of actors and accelerating long-run wealth accumulation and inequality (Skare et al. 2024).<br>• LLMs may compress high wages (reducing 90:10 gaps) while increasing top-tail inequality as returns concentrate among elite talent and capital (Peppiatt 2024).<br>• Unequal access and widening digital divides in connectivity, compute, and AI literacy may deepen global disparities (Hagerty and Rubinov 2019; Mirza et al. 2024; Mohamed et al. 2020; Vargas 2025).<br>• Data-ownership asymmetries enable firms to extract and monetise user-generated data without compensation, allowing value capture from uncompensated inputs and widening economic inequality (Epstein et al. 2023; Mulligan and Godsiff 2023; Precel et al. 2024).<br>• Data and algorithmic bias produce both allocative and representational harms, reinforcing structural inequities (An et al. 2025; Capraro et al. 2024; Hofmann et al. 2024; Lloyd 2018; Neumann et al. 2024).<br>• Data colonialism and extractive AI value-chains may appropriate data, labour, and ecological resources from Global South communities without adequate compensation, potentially entrenching Global North–South asymmetries (Hagerty and Rubinov 2019; Mirza et al. 2024; Mohamed et al. 2020; Vargas 2025).</td><td>• Limited research on AI-led wealth concentration.<br>• Understudied Global North–South divide in AI.<br>• Lack of intersectional analysis in AI fairness.<br>• Scarcity of participatory design approaches involving marginalised and underrepresented groups.</td></tr>
<tr><td>Employment</td><td>• AI is affecting work primarily at the task level, through automation and augmentation, with documented displacement risks and heterogeneous productivity effects (Acemoglu 2024a; Acemoglu and Restrepo 2019; Ozgul et al. 2024; Qiao et al. 2026).<br>• Occupations tend to exhibit an inflection point where complementarity persists until AI surpasses task complexity, after which displacement occurs (Qiao et al. 2026).</td><td>• Lack of granular, longitudinal evidence on real labour market outcomes.<br>• Poor understanding of substitution vs. complementarity dynamics.<br>• Limited analysis of global and cross-border effects.</td></tr>
</table>

| Impact type | Full findings | Research gaps |
| --- | --- | --- |
| | • High-skilled, knowledge-intensive jobs (e.g. programming, writing, finance, law) are most exposed to LLMs (Eloundou et al. 2024; Felten et al. 2021).<br>• Senior roles retain a human premium due to lower substitutability (task diversity, client exposure, and personal reputation), while junior and entry-level roles are more exposed to displacement (Xie and Avila 2025).<br>• Productivity gains are concentrated among novice and lower-performing workers, narrowing performance gaps (Noy and Zhang 2023).<br>• AI contributes to a declining labour income share, eroding worker bargaining power (Acemoglu 2024a; Acemoglu and Restrepo 2018; Acemoglu et al. 2022; Aghion et al. 2019; Autor and Salomons 2018; Hubmer and Restrepo 2021; Lu and Zhou 2021; Peppiatt 2024; Trammell and Korinek 2023).<br>• Gig work and microwork underpin AI development but are low-paid, precarious and mostly performed by women and workers in the Global South, with the risk of increasing job insecurity and inequality (Altenried 2020; Mohla et al. 2024; Nwachukwu et al. 2023; Vargas 2025).<br>• Algorithmic management tends to intensify worker surveillance and standardisation, reducing autonomy and increasing psychosocial stress (Deranty and Corbin 2022).<br>• Bias in hiring and human resources management tools may perpetuate gender and ethnic disparities (Ferrara 2024). | • Impacts on job quality (job satisfaction, autonomy, meaning, psychosocial impacts) are underexplored.<br>• Lack of analysis of the impacts of AI on informal work. |
| ***Health*** | • AI may improve diagnosis, treatment, and surgeries and is most effective when complementing human care (Saccà et al. 2023).<br>• Benefits are unequally distributed, exacerbating health inequities via bias and unequal access (Paik et al. 2023; Rinderknecht et al. 2024; Van Kolfschooten 2023).<br>• AI systems may expand health and mental-health support but pose risks of overreliance (Alabed et al. 2024).<br>• Privacy, informed-consent, and liability gaps create governance and safety concerns (Mohammad Amini et al. 2023; Rosic 2024; Saadat et al. 2024). | • Limited empirical evidence on AI's clinical and equity outcomes.<br>• Insufficient regulatory and ethical frameworks for fairness, informed consent, and liability in AI for health. |
| ***Growth and income*** | • Long-run and theoretical growth projections span the full spectrum, from negligible to potentially unbounded, depending on model assumptions (Trammell and Korinek 2023; Wynne and Derr 2025).<br>• By contrast, mainstream economic forecasts predict modest near-term gains from AI adoption, with total factor productivity rising by only 0.53–0.71% and GDP by 0.93–1.56% over 10 years (Acemoglu 2024b).<br>• Firm-level productivity gains are concentrated in marketing, customer service, media, software development, and data-intensive sectors, and depend on complementary investments (Acemoglu 2024b).<br>• Early AI deployments may reconfigure production, yielding capital deepening and productivity gains in specific sectors, with system-level substitution and scale feedback delaying economy-wide productivity gains (Brynjolfsson et al. 2017).<br>• Transformative AI may trigger rapid or potentially exponential output growth while collapsing labour share, even if wages rise, due to capital deepening (Trammell and Korinek 2023).<br>• Income distribution systems that are decoupled from labour earnings may be needed as AI reduces labour's role in production and income generation (Korinek and Juelfs 2022; Trammell and Korinek 2023).<br>• Some AI-generated tasks (e.g. deepfakes, manipulative ads) may increase GDP (e.g. +2%) while reducing welfare (e.g. −0.72%; Acemoglu 2024b). | • Limited empirical validation of macroeconomic models.<br>• High uncertainty on the projected impact of AI on GDP growth.<br>• Ambiguity in AI modelling and measurement, with theoretical models treating AI broadly and empirical studies examining narrow applications.<br>• Scarcity of firm-level AI adoption data, limiting analysis of heterogeneous sectoral, regional, and firm-size effects. |
| ***Existential risk*** | • Misaligned or insufficiently controllable advanced AI systems may behave in unpredictable, power-seeking, or goal-divergent ways, creating loss-of-control scenarios that threaten long-term human survival (Bengio et al. 2024; Goldfarb 2024; Tamburrini 2024; Turchin and Denkenberger 2020).<br>• Emergence of superior nonhuman intelligence might displace human agency and influence, threatening long-term species survival (Bengio et al. 2024; Goldfarb 2024; Tamburrini 2024; Turchin and Denkenberger 2020).<br>• Jailbreaking and malicious use of AI models might enable bioweapon design, cyberattacks, and other catastrophic misuse (Bengio et al. 2024; Goldfarb 2024; Krishna 2024; Tamburrini 2024; Turchin and Denkenberger 2020).<br>• Autonomous weapons and AI-based warfare might escalate conflicts without human oversight, raising the likelihood of large-scale warfare (Bengio et al. 2024; Benjamins and Salazar 2020; Goldfarb 2024; Tamburrini 2024; Turchin and Denkenberger 2020).<br>• AI-generated deepfakes and autonomous systems might undermine nuclear deterrence by eroding second-strike credibility and enabling deception in crisis scenarios (Tamburrini 2024).<br>• Proposals to integrate AI into nuclear command, control, and early-warning systems introduce risks of automation errors, adversarial manipulation, and reduced human oversight, potentially accelerating inadvertent escalation or mistaken launch decisions (Tamburrini 2024). | • Limited research on alignment and control strategies for highly capable AI systems.<br>• Limited analysis of adversarial vulnerabilities, automation errors, and human-machine interaction risks in nuclear command, control, and early-warning systems.<br>• Scarce research on verifiable governance mechanisms for restricting high-risk AI development, testing, and deployment on a global scale.<br>• Lack of probabilistic risk estimates and early-warning indicators. |

| Impact type | Full findings | Research gaps |
|---|---|---|
| ***Subjective well-being*** | • Higher AI investment is associated with lower current and expected subjective well-being across 137 countries (Zhao et al. 2024).<br>• Negative effects concentrate among youth, men, high-income and high-skilled workers, and manufacturing employees (Zhao et al. 2024).<br>• Well-being impacts are likely mediated by factors such as anxiety over job loss, income instability, privacy concerns, and perceived governance gaps (Zhao et al. 2024).<br>• AI can improve access to mental health support, emotional companionship, and personalised care, but may also foster overdependence, reduce autonomy and empathy, or contribute to social isolation (Alabed et al. 2024; Gupta et al. 2023; Kieslich et al. 2024; Ma et al. 2023; Marriott and Pitardi 2024).<br>• AI can either improve or harm workers' well-being by boosting productivity and satisfaction on the one hand, or by increasing stress, burnout, and identity threat on the other (Peppiatt 2024). | • Research on AI's impact on subjective well-being is scarce and predominantly conceptual, with limited empirical evidence.<br>• Limited research on causal mechanisms linking AI exposure to subjective well-being.<br>• Lack of longitudinal evidence tracking well-being effects over time.<br>• Lack of research on multidimensional well-being and time use. |
| ***Cognitive capabilities*** | • AI-mediated personalised learning may improve educational outcomes (Qasmi and Fatima 2024).<br>• Extended exposure to LLMs may facilitate cognitive offloading, impairing deep engagement and active learning (Dergaa et al. 2024; Gerlich 2025; Kosmyna et al. 2025).<br>• Reliance on AI assistance may weaken memory encoding and perceived ownership of written content (Dergaa et al. 2024; Gerlich 2025; Kosmyna et al. 2025).<br>• AI-assisted writing reduces activation in brain regions linked to creativity, semantic integration, and executive control (Dergaa et al. 2024; Gerlich 2025; Kosmyna et al. 2025).<br>• LLM use may promote linguistic and conceptual homogenisation, limiting diversity of thought and expression (Dergaa et al. 2024; Gerlich 2025; Kosmyna et al. 2025).<br>• The prolonged reliance on LLMs might lead to the accumulation of cognitive debt, diminishing critical thinking and problem-solving skills (Dergaa et al. 2024; Gerlich 2025; Kosmyna et al. 2025).<br>• Younger users and those with lower educational attainment are more susceptible to reduced critical thinking due to AI reliance (Dergaa et al. 2024; Gerlich 2025; Kosmyna et al. 2025). | • Long-term effects like cognitive debt and autonomy loss are poorly understood.<br>• Lack of causal and longitudinal evidence, and analyses of heterogeneity and mediators.<br>• Sparse and uneven evidence across populations, real-world contexts, and cognitive domains.<br>• Limited understanding of tool- and design-specific effects across AI systems and interfaces. |

## 3.2 Notes on the tables in the main text

Tables 1 and 2 in the main text have been derived from the expanded tables presented in Section 3.1. After compiling the full tables of findings and research gaps, each item was evaluated along two dimensions: (i) the severity of the impact, assessed as high or low based on the expected significance of the effect; and (ii) the degree of confidence, assessed from the number of studies supporting the item and the confidence level assigned to those studies. For each level and type of impact, the tables in the main text include the items classified as having the highest impact and the highest level of confidence. This procedure ensured that the tables in the article highlight the most robust and policy-relevant findings and research gaps.

## 3.3 Notes on the unified framework for AI's environmental and well-being impacts

Fig. 5, "Unified framework for assessing the environmental and well-being impacts of AI", shows an illustrative conceptual framework organising how AI may affect environmental sustainability and human well-being across three categories. The framework is adapted from the three-level structure proposed by Kaack et al. (2022) for greenhouse gas (GHG) impacts, extended here to include broader environmental sustainability impacts and well-being impacts.

The categories considered are:

- A. Computing-related impacts;
- B. Application-level impacts;
- C. Systemic impacts.

See Section 1.7.2 for definitions of these three categories.

The upper part of the figure shows impacts on human well-being, while the lower part shows impacts on environmental sustainability. Application-level impacts in B are organised around sectoral domains. For the environmental side, these domains are adapted from the sectors included in the original framework by Kaack et al. (2022), with some sectors merged or broadened to fit the unified framework. For the well-being side, the sectoral structure is expanded to include domains where the literature identifies direct social impacts, such as healthcare, education and research, information and communication, industry and consumer services, and infrastructure and the built environment. Some sector boxes are positioned across the environmental and well-being parts of the figure to indicate that the same AI applications can generate impacts in both dimensions. Supplementary Table 3 shows how the environmental sectors from the original framework proposed by Kaack et al. (2022) map onto the environmental part of the unified framework.

**Supplementary Table 3 Mapping of environmental application-level sectors from the original framework to the unified framework.** The table shows how the environmental sectors in the framework proposed by Kaack et al. (2022) were retained, merged, or broadened in the environmental part of the unified framework

| Original framework proposed by Kaack et al. (2022) | Unified framework |
| --- | --- |
| • Industry | • Industry & consumer services |
| • Buildings<br>• Transportation | • Infrastructure & built environment |
| • Power systems<br>• Energy supply | • Power systems & energy supply |
| • Agriculture | • Agriculture |
| • Forestry and other land use | • Forestry and other land use |

Arrows depict likely directions of change, not magnitude or certainty. Green arrows and labels indicate impacts likely to increase environmental sustainability or human well-being, while magenta arrows and labels indicate impacts likely to decrease environmental sustainability or human well-being under typical conditions. No inference should be drawn about which tendency is likely to prevail in any given sector or category.

The schematic is illustrative, not exhaustive. It serves as a conceptual guide to organise findings across computing-related, application-level, and systemic impacts, and to highlight trade-offs, research gaps, and overlooked interactions between environmental sustainability and human well-being.

## References

Acemoglu D (2024a) Harms of AI. In: Bullock JB, Chen Y-C, Himmelreich J, et al. (eds) The Oxford Handbook of AI Governance. Oxford University Press, pp 660–706. https://doi.org/10.1093/oxfordhb/9780197579329.013.65

Acemoglu D (2024b) The Simple Macroeconomics of AI. NBER Working Paper 32487. National Bureau of Economic Research. https://doi.org/10.3386/w32487

Acemoglu D, Anderson GW, Beede DN, et al (2022) Automation and the Workforce: A Firm-Level View from the 2019 Annual Business Survey. NBER Working Paper 30659. National Bureau of Economic Research. https://doi.org/10.3386/w30659

Acemoglu D, Restrepo P (2018) The Race between Man and Machine: Implications of Technology for Growth, Factor Shares, and Employment. American Economic Review 108:1488–1542. https://doi.org/10.1257/aer.20160696

Acemoglu D, Restrepo P (2019) Artificial Intelligence, Automation, and Work. In: Agrawal A, Gans J, Goldfarb A (eds) The Economics of Artificial Intelligence: An Agenda. University of Chicago Press, pp 197–236. https://doi.org/10.7208/chicago/9780226613475.003.0008

Aghion P, Antonin C, Bunel S (2019) Artificial intelligence, growth and employment: The role of policy. Économie et Statistique 510–511–512:149–164. https://doi.org/10.24187/ecostat.2019.510t.1994

Alabed A, Javornik A, Gregory-Smith D, Casey R (2024) More than just a chat: a taxonomy of consumers' relationships with conversational AI agents and their well-being implications. European Journal of Marketing 58:373–409. https://doi.org/10.1108/EJM-01-2023-0037

Alswaitti M, Verdecchia R, Danoy G, et al (2024) Training Green AI Models Using Elite Samples. Preprint at https://doi.org/10.48550/arXiv.2402.12010

Altenried M (2020) The platform as factory: Crowdwork and the hidden labour behind artificial intelligence. Capital & Class 44:145–158. https://doi.org/10.1177/0309816819899410

An J, Huang D, Lin C, Tai M (2025) Measuring gender and racial biases in large language models: Intersectional evidence from automated resume evaluation. PNAS Nexus 4:pgaf089. https://doi.org/10.1093/pnasnexus/pgaf089

Autor D, Salomons A (2018) Is Automation Labor Share-Displacing? Productivity Growth, Employment, and the Labor Share. Brookings Papers on Economic Activity 2018:1–87. https://doi.org/10.1353/eca.2018.0000

Bastos J, Monforti-Ferrario F, Melica G (2024) GHG Emission Factors for Electricity Consumption. JRC Technical Report JRC136340. European Commission, Joint Research Centre. https://publications.jrc.ec.europa.eu/repository/handle/JRC136340

Bender EM, Gebru T, McMillan-Major A, Shmitchell S (2021) On the Dangers of Stochastic Parrots: Can Language Models Be Too Big? 🦜. In: Proceedings of the 2021 ACM Conference on Fairness, Accountability, and Transparency. Association for Computing Machinery, New York, NY, USA, pp 610–623. https://doi.org/10.1145/3442188.3445922

Bengio Y, Hinton G, Yao A, et al (2024) Managing extreme AI risks amid rapid progress. Science 384:842–845. https://doi.org/10.1126/science.adn0117

Benjamins R, Salazar I (2020) Towards a framework for understanding societal and ethical implications of Artificial Intelligence. Preprint at https://doi.org/10.48550/arXiv.2001.09750

Birhane A (2022) Automating Ambiguity: Challenges and Pitfalls of Artificial Intelligence. Preprint at https://doi.org/10.48550/arXiv.2206.04179

Brynjolfsson E, Rock D, Syverson C (2017) Artificial Intelligence and the Modern Productivity Paradox: A Clash of Expectations and Statistics. NBER Working Paper 24001. National Bureau of Economic Research. https://doi.org/10.3386/w24001

Capraro V, Lentsch A, Acemoglu D, et al (2024) The impact of generative artificial intelligence on socioeconomic inequalities and policy making. PNAS Nexus 3:pgae191. https://doi.org/10.1093/pnasnexus/pgae191

Carmeno NL, Domingos T, O'Neill DW (2026) Systematic Literature Review Protocol: The impacts of artificial intelligence on environmental sustainability and human well-being. protocols.io. https://doi.org/10.17504/protocols.io.j8nlkzjwwl5r/v1

Chen C-T, Khan A, Chen S-C (2024) Modeling the impact of BDA-AI on sustainable innovation ambidexterity and environmental performance. Journal of Big Data 11:124. https://doi.org/10.1186/s40537-024-00995-6

Chien AA, Lin L, Nguyen H, et al (2023) Reducing the Carbon Impact of Generative AI Inference (today and in 2035). In: Proceedings of the 2nd Workshop on Sustainable Computer Systems. ACM, Boston, MA, USA, pp 1–7. https://doi.org/10.1145/3604930.3605705

Coveri A, Cozza C, Guarascio D (2025) Big Tech and the US Digital-Military-Industrial Complex. Intereconomics 60:81–87. https://doi.org/10.2478/ie-2025-0017

de Vries A (2023) The growing energy footprint of artificial intelligence. Joule 7:2191–2194. https://doi.org/10.1016/j.joule.2023.09.004

Deranty J-P, Corbin T (2022) Artificial Intelligence and work: a critical review of recent research from the social sciences. AI & Society 39:675–691. https://doi.org/10.1007/s00146-022-01496-x

Dergaa I, Ben Saad H, Glenn JM, et al (2024) From tools to threats: a reflection on the impact of artificial-intelligence chatbots on cognitive health. Frontiers in Psychology 15:1259845. https://doi.org/10.3389/fpsyg.2024.1259845

Dodge J, Prewitt T, Tachet Des Combes R, et al (2022) Measuring the Carbon Intensity of AI in Cloud Instances. In: Proceedings of the 2022 ACM Conference on Fairness, Accountability, and Transparency. Association for Computing Machinery, New York, NY, USA, pp 1877–1894. https://doi.org/10.1145/3531146.3533234

Eloundou T, Manning S, Mishkin P, Rock D (2024) GPTs are GPTs: Labor market impact potential of LLMs. Science 384:1306–1308. https://doi.org/10.1126/science.adj0998

Epstein Z, Hertzmann A, Herman L, et al (2023) Art and the science of generative AI: A deeper dive. Science 380:1110–1111. https://doi.org/10.1126/science.adh4451

Falk S, Ekchajzer D, Pirson T, et al (2025) More than Carbon: Cradle-to-Grave environmental impacts of GenAI training on the Nvidia A100 GPU. Preprint at https://doi.org/10.48550/arXiv.2509.00093

Felten E, Raj M, Seamans R (2021) Occupational, industry, and geographic exposure to artificial intelligence: A novel dataset and its potential uses. Strategic Management Journal 42:2195–2217. https://doi.org/10.1002/smj.3286

Ferrara E (2024) The Butterfly Effect in Artificial Intelligence Systems: Implications for AI Bias and Fairness. Machine Learning with Applications 15:100525. https://doi.org/10.1016/j.mlwa.2024.100525

Forti V, Baldé CP, Kuehr R, Bel G (2020) The Global E-waste Monitor 2020: Quantities, flows, and the circular economy potential. United Nations University/United Nations Institute for Training and Research, International Telecommunication Union, and International Solid Waste Association, Bonn, Geneva, and Rotterdam. https://ewastemonitor.info/gem-2020/

Frost K, Hua I (2019) Quantifying spatiotemporal impacts of the interaction of water scarcity and water use by the global semiconductor manufacturing industry. Water Resources and Industry 22:100115. https://doi.org/10.1016/j.wri.2019.100115

Fu Y, Shen Y, Song M, Wang W (2024) Does artificial intelligence reduce corporate energy consumption? New evidence from China. Economic Analysis and Policy 83:548–561. https://doi.org/10.1016/j.eap.2024.07.005

Gaffney O, Luers A, Carrero-Martinez F, et al (2025) The Earth alignment principle for artificial intelligence. Nature Sustainability 8:467–469. https://doi.org/10.1038/s41893-025-01536-6

Gerlich M (2025) AI Tools in Society: Impacts on Cognitive Offloading and the Future of Critical Thinking. Societies 15:6. https://doi.org/10.3390/soc15010006

Ghobakhloo M, Fathi M, Iranmanesh M, et al (2024) Generative artificial intelligence in manufacturing: opportunities for actualizing Industry 5.0 sustainability goals. Journal of Manufacturing Technology Management 35:94–121. https://doi.org/10.1108/JMTM-12-2023-0530

Gibney E (2022) How to shrink AI's ballooning carbon footprint. Nature 607:648. https://doi.org/10.1038/d41586-022-01983-7

Goldfarb A (2024) Pause artificial intelligence research? Understanding AI policy challenges. Canadian Journal of Economics 57:363–377. https://doi.org/10.1111/caje.12705

Gossen M, Lell O (2023) Sustainable consumption in the digital age: A plea for a systemic policy approach to turn risks into opportunities. GAIA - Ecological Perspectives for Science and Society 32:71–76. https://doi.org/10.14512/gaia.32.S1.11

Guidi G, Dominici F, Gilmour J, et al (2024) Environmental Burden of United States Data Centers in the Artificial Intelligence Era. Preprint at https://doi.org/10.48550/arXiv.2411.09786

Gupta D, Singhal A, Sharma S, et al (2023) Humans' Emotional and Mental Well-Being under the Influence of Artificial Intelligence. Journal for ReAttach Therapy and Developmental Diversities 6:184–197. https://doi.org/10.67145/jrtdd.v6i6s.698

Hagerty A, Rubinov I (2019) Global AI Ethics: A Review of the Social Impacts and Ethical Implications of Artificial Intelligence. Preprint at https://doi.org/10.48550/arXiv.1907.07892

Hofmann V, Kalluri PR, Jurafsky D, King S (2024) AI generates covertly racist decisions about people based on their dialect. Nature 633:147–154. https://doi.org/10.1038/s41586-024-07856-5

Hubmer J, Restrepo P (2021) Not a Typical Firm: The Joint Dynamics of Firms, Labor Shares, and Capital–Labor Substitution. NBER Working Paper 28579. National Bureau of Economic Research. https://doi.org/10.3386/w28579

Hussein BR, Halimu C, Siddique MT (2021) The Future of Artificial Intelligence and its Social, Economic and Ethical Consequences. Preprint at https://doi.org/10.48550/arXiv.2101.03366

Jagannadharao A, Beckage N, Nafus D, Chamberlin S (2023) Timeshifting strategies for carbon-efficient long-running large language model training. Innovations in Systems and Software Engineering. https://doi.org/10.1007/s11334-023-00546-x

Jiang P, Sonne C, Li W, et al (2024) Preventing the Immense Increase in the Life-Cycle Energy and Carbon Footprints of LLM-Powered Intelligent Chatbots. Engineering 40:202–210. https://doi.org/10.1016/j.eng.2024.04.002

Kaack LH, Donti PL, Strubell E, et al (2022) Aligning artificial intelligence with climate change mitigation. Nature Climate Change 12:518–527. https://doi.org/10.1038/s41558-022-01377-7

Kalluri PR, Agnew W, Cheng M, et al (2023) The Surveillance AI Pipeline. Preprint at https://doi.org/10.48550/arXiv.2309.15084

Kamiya G, Coroamă VC (2025) Data Centre Energy Use: Critical Review of Models and Results. EDNA–IEA 4E TCP. https://www.iea-4e.org/edna/publications/data-centre-energy-use-critical-review-of-models-and-results/

Kieslich K, Helberger N, Diakopoulos N (2024) My Future with My Chatbot: A Scenario-Driven, User-Centric Approach to Anticipating AI Impacts. In: Proceedings of the 2024 ACM Conference on Fairness, Accountability, and Transparency. Association for Computing Machinery, New York, NY, USA, pp 2071–2085. https://doi.org/10.1145/3630106.3659026

Korinek A, Juelfs M (2022) Preparing for the (Non-Existent?) Future of Work. NBER Working Paper 30172. National Bureau of Economic Research. https://doi.org/10.3386/w30172

Kosmyna N, Hauptmann E, Yuan YT, et al (2025) Your Brain on ChatGPT: Accumulation of Cognitive Debt when Using an AI Assistant for Essay Writing Task. Preprint at https://doi.org/10.48550/arXiv.2506.08872

Krishna VV (2024) AI and contemporary challenges: The good, bad and the scary. Journal of Open Innovation: Technology, Market, and Complexity 10:100178. https://doi.org/10.1016/j.joitmc.2023.100178

Kshetri N (2024) The Environmental Impact of Artificial Intelligence. IT Professional 26:9–13. https://doi.org/10.1109/MITP.2024.3399471

Lehuedé S (2025) An elemental ethics for artificial intelligence: water as resistance within AI's value chain. AI & Society 40:1761–1774. https://doi.org/10.1007/s00146-024-01922-2

Li EC-Y, Pang K-W (2024) Fandom meets artificial intelligence: Rethinking participatory culture as human–community–machine interactions. European Journal of Cultural Studies 27:778–787. https://doi.org/10.1177/13675494241236146

Li P, Yang J, Islam MA, Ren S (2025) Making AI Less "Thirsty": Uncovering and Addressing the Secret Water Footprint of AI Models. Communications of the ACM 68:54–63. https://doi.org/10.1145/3724499

Li X, Wang Q, Tang Y (2024) The Impact of Artificial Intelligence Development on Urban Energy Efficiency—Based on the Perspective of Smart City Policy. Sustainability 16:3200. https://doi.org/10.3390/su16083200

Lin D, Hanscom L, Murthy A, et al (2018) Ecological Footprint Accounting for Countries: Updates and Results of the National Footprint Accounts, 2012–2018. Resources 7:58. https://doi.org/10.3390/resources7030058

Liu J, Qian Y, Yang Y, Yang Z (2022) Can Artificial Intelligence Improve the Energy Efficiency of Manufacturing Companies? Evidence from China. International Journal of Environmental Research and Public Health 19:2091. https://doi.org/10.3390/ijerph19042091

Lloyd K (2018) Bias Amplification in Artificial Intelligence Systems. Preprint at https://doi.org/10.48550/arXiv.1809.07842

Lu Y, Zhou Y (2021) A review on the economics of artificial intelligence. Journal of Economic Surveys 35:1045–1072. https://doi.org/10.1111/joes.12422

Luccioni AS, Gamazaychikov B, Costa TA da, Strubell E (2025a) Misinformation by Omission: The Need for More Environmental Transparency in AI. Preprint at https://doi.org/10.48550/arXiv.2506.15572

Luccioni AS, Hernandez-Garcia A (2023) Counting Carbon: A Survey of Factors Influencing the Emissions of Machine Learning. Preprint at https://doi.org/10.48550/arXiv.2302.08476

Luccioni AS, Strubell E, Crawford K (2025b) From Efficiency Gains to Rebound Effects: The Problem of Jevons' Paradox in AI's Polarized Environmental Debate. In: Proceedings of the 2025 ACM Conference on Fairness, Accountability, and Transparency. Association for Computing Machinery, New York, NY, USA, pp 76–88. https://doi.org/10.1145/3715275.3732007

Ma Z, Mei Y, Su Z (2023) Understanding the Benefits and Challenges of Using Large Language Model-based Conversational Agents for Mental Well-being Support. AMIA Annual Symposium Proceedings 2023:1105–1114.

Marriott HR, Pitardi V (2024) One is the loneliest number… Two can be as bad as one. The influence of AI Friendship Apps on users' well-being and addiction. Psychology and Marketing 41:86–101. https://doi.org/10.1002/mar.21899

McKinney S (2024) Integrating Artificial Intelligence into Citizens' Assemblies: Benefits, Concerns and Future Pathways. Journal of Deliberative Democracy 20:1556. https://doi.org/10.16997/jdd.1556

Mirza S, Coelho B, Cui Y, et al (2024) Global-Liar: Factuality of LLMs over Time and Geographic Regions. Preprint at https://doi.org/10.48550/arXiv.2401.17839

Mitra B, Cramer H, Gurevich O (2025) Sociotechnical Implications of Generative Artificial Intelligence for Information Access. In: White RW, Shah C (eds) Information Access in the Era of Generative AI. Springer Nature Switzerland, Cham, pp 161–200. https://doi.org/10.1007/978-3-031-73147-1_7

Mohamed S, Png M-T, Isaac W (2020) Decolonial AI: Decolonial Theory as Sociotechnical Foresight in Artificial Intelligence. Philosophy & Technology 33:659–684. https://doi.org/10.1007/s13347-020-00405-8

Mohammad Amini M, Jesus M, Fanaei Sheikholeslami D, et al (2023) Artificial Intelligence Ethics and Challenges in Healthcare Applications: A Comprehensive Review in the Context of the European GDPR Mandate. Machine Learning and Knowledge Extraction 5:1023–1035. https://doi.org/10.3390/make5030053

Mohla S, Bagh B, Guha A (2024) Thinking beyond Bias: Analyzing Multifaceted Impacts and Implications of AI on Gendered Labour. Preprint at https://doi.org/10.48550/arXiv.2406.16207

Mulligan CEA, Godsiff P (2023) Datalism and Data Monopolies in the Era of A.I.: A Research Agenda. Preprint at https://doi.org/10.48550/arXiv.2307.08049

Natarajan HK, de Paula D, Dremel C, Uebernickel F (2022) A Theoretical Review on AI Affordances for Sustainability. In: 28th Americas Conference on Information Systems (AMCIS 2022). Association for Information Systems. https://aisel.aisnet.org/amcis2022/sig_green/sig_green/13/

Neumann T, Lee S, De-Arteaga M, et al (2024) Diverse, but Divisive: LLMs Can Exaggerate Gender Differences in Opinion Related to Harms of Misinformation. Preprint at https://doi.org/10.48550/arXiv.2401.16558

Niu C, Zhang W, Li J, et al (2025) TokenPowerBench: Benchmarking the Power Consumption of LLM Inference. Preprint at https://doi.org/10.48550/arXiv.2512.03024

Noy S, Zhang W (2023) Experimental evidence on the productivity effects of generative artificial intelligence. Science 381:187–192. https://doi.org/10.1126/science.adh2586

Nwachukwu NM, Roberts JS, Montoya LN (2023) The Glamorisation of Unpaid Labour: AI and its Influencers. Preprint at https://doi.org/10.48550/arXiv.2308.02399

OpenAI (2025) What are tokens and how to count them? OpenAI Help Center. https://help.openai.com/en/articles/4936856-what-are-tokens-and-how-to-count-them. Accessed 16 Jan 2026

Ozgul P, Fregin M-C, Stops M, et al (2024) High-skilled Human Workers in Non-Routine Jobs are Susceptible to AI Automation but Wage Benefits Differ between Occupations. Preprint at https://doi.org/10.48550/arXiv.2404.06472

Page MJ, McKenzie JE, Bossuyt PM, et al (2021) The PRISMA 2020 statement: An updated guideline for reporting systematic reviews. PLOS Medicine 18:e1003583. https://doi.org/10.1371/journal.pmed.1003583

Paik KE, Hicklen R, Kaggwa F, et al (2023) Digital Determinants of Health: Health data poverty amplifies existing health disparities—A scoping review. PLOS Digital Health 2:e0000313. https://doi.org/10.1371/journal.pdig.0000313

Pappalardo L, Ferragina E, Citraro S, et al (2024) A survey on the impact of AI-based recommenders on human behaviours: methodologies, outcomes and future directions. Preprint at https://doi.org/10.48550/arXiv.2407.01630

Peppiatt C (2024) The Future of Work: Inequality, Artificial Intelligence, and What Can Be Done About It. A Literature Review. Preprint at https://doi.org/10.48550/arXiv.2408.13300

Precel H, McDonald A, Hecht B, Vincent N (2024) A Canary in the AI Coal Mine: American Jews May Be Disproportionately Harmed by Intellectual Property Dispossession in Large Language Model Training. In: Proceedings of the CHI Conference on Human Factors in Computing Systems. Association for Computing Machinery, New York, NY, USA, pp 1–17. https://doi.org/10.1145/3613904.3642749

Qasmi F, Fatima SK (2024) Enhancing Education and Well-Being Through Artificial Intelligence: Opportunities and Challenges. Communications in Computer and Information Science 2162 CCIS:154–164. https://doi.org/10.1007/978-3-031-65996-6_13

Qiao D, Rui H, Xiong Q (2026) Artificial Intelligence and Jobs: Has the Inflection Point Arrived? Evidence from an Online Labor Platform. Journal of Management Information Systems 43:5–37. https://doi.org/10.1080/07421222.2025.2602400

Rikap C (2024) Varieties of corporate innovation systems and their interplay with global and national systems: Amazon, Facebook, Google and Microsoft's strategies to produce and appropriate artificial intelligence. Review of International Political Economy 31:1735–1763. https://doi.org/10.1080/09692290.2024.2365757

Rinderknecht F-A, Nwandu L, Daneshjou R, Lester J (2024) The Impact of Artificial Intelligence on Health Equity in Dermatology. Current Dermatology Reports 13:148–155. https://doi.org/10.1007/s13671-024-00436-w

Rosic A (2024) Legal implications of artificial intelligence in health care. Clinics in Dermatology 42:451–459. https://doi.org/10.1016/j.clindermatol.2024.06.014

Saadat A, Siddiqui T, Taseen S, Mughal S (2024) Revolutionising Impacts of Artificial Intelligence on Health Care System and Its Related Medical In-Transparencies. Annals of Biomedical Engineering 52:1546–1548. https://doi.org/10.1007/s10439-023-03343-6

Saccà R, Turrini R, Ausania F, et al (2023) The ménage à trois of healthcare: the actors in after-AI era under patient consent. Frontiers in Medicine 10:1329087. https://doi.org/10.3389/fmed.2023.1329087

Shoaib MR, Wang Z, Ahvanooey MT, Zhao J (2023) Deepfakes, Misinformation, and Disinformation in the Era of Frontier AI, Generative AI, and Large AI Models. In: 2023 International Conference on Computer and Applications (ICCA). pp 1–7. https://doi.org/10.1109/ICCA59364.2023.10401723

Skare M, Gavurova B, Blažević Burić S (2024) Artificial intelligence and wealth inequality: A comprehensive empirical exploration of socioeconomic implications. Technology in Society 79:102719. https://doi.org/10.1016/j.techsoc.2024.102719

Tamburrini G (2024) Artificial Intelligence and Large-Scale Threats to Humanity. In: Werthner H, Ghezzi C, Kramer J, et al. (eds) Introduction to Digital Humanism. Springer Nature Switzerland, Cham, pp 241–254. https://doi.org/10.1007/978-3-031-45304-5_16

Tessler MH, Bakker MA, Jarrett D, et al (2024) AI can help humans find common ground in democratic deliberation. Science 386:eadq2852. https://doi.org/10.1126/science.adq2852

Trammell P, Korinek A (2023) Economic Growth under Transformative AI. NBER Working Paper 31815. National Bureau of Economic Research. https://doi.org/10.3386/w31815

Turchin A, Denkenberger D (2020) Classification of global catastrophic risks connected with artificial intelligence. AI & Society 35:147–163. https://doi.org/10.1007/s00146-018-0845-5

Van Kolfschooten H (2023) The AI cycle of health inequity and digital ageism: mitigating biases through the EU regulatory framework on medical devices. Journal of Law and the Biosciences 10:lsad031. https://doi.org/10.1093/jlb/lsad031

Vargas NC (2025) Exploiting the Margin: How Capitalism Fuels AI at the Expense of Minoritized Groups. AI and Ethics 5:1871–1876. https://doi.org/10.1007/s43681-024-00502-w

Varoquaux G, Luccioni AS, Whittaker M (2025) Hype, Sustainability, and the Price of the Bigger-is-Better Paradigm in AI. In: Proceedings of the 2025 ACM Conference on Fairness, Accountability, and Transparency. Association for Computing Machinery, New York, NY, USA, pp 61–75. https://doi.org/10.1145/3715275.3732006

Verdecchia R, Cruz L, Sallou J, et al (2022) Data-Centric Green AI: An Exploratory Empirical Study. In: 2022 International Conference on ICT for Sustainability (ICT4S). IEEE, pp 35–45. https://doi.org/10.1109/ICT4S55073.2022.00015

Verdegem P (2024) Dismantling AI capitalism: the commons as an alternative to the power concentration of Big Tech. AI & Society 39:727–737. https://doi.org/10.1007/s00146-022-01437-8

Wang J, Wang A, Luo K, Nie Y (2024a) Can artificial intelligence improve enterprise environmental performance: Evidence from China. Journal of Environmental Management 370:123079. https://doi.org/10.1016/j.jenvman.2024.123079

Wang P, Zhang L-Y, Tzachor A, Chen W-Q (2024b) E-waste challenges of generative artificial intelligence. Nature Computational Science 4:818–823. https://doi.org/10.1038/s43588-024-00712-6

Wang Q, Li Y, Li R (2024c) Ecological footprints, carbon emissions, and energy transitions: the impact of artificial intelligence (AI). Humanities and Social Sciences Communications 11:1–18. https://doi.org/10.1057/s41599-024-03520-5

Willenbacher M (2021) A Short Overview of Rebound Effects in Methods of Artificial Intelligence. International Journal of Environmental Sciences & Natural Resources 28:556246. https://doi.org/10.19080/IJESNR.2021.28.556246

Wright D, Igel C, Samuel G, Selvan R (2025) Efficiency Is Not Enough: A Critical Perspective on Environmentally Sustainable AI. Communications of the ACM 68:62–69. https://doi.org/10.1145/3724500

Wu C-J, Raghavendra R, Gupta U, et al (2022) Sustainable AI: Environmental Implications, Challenges and Opportunities. Proceedings of Machine Learning and Systems 4:795–813.

Wynne MA, Derr L (2025) Advances in AI will boost productivity, living standards over time. Dallas Fed Economics. https://www.dallasfed.org/research/economics/2025/0624. Accessed 16 Dec 2025

Xie Y, Avila S (2025) The social impact of generative LLM-based AI. Chinese Journal of Sociology 11:31–57. https://doi.org/10.1177/2057150X251315997

Zhang G, Chang F, Liu J (2024) Carbon emission prediction of 275 cities in China considering artificial intelligence effects and feature interaction: A heterogeneous deep learning modeling framework. Sustainable Cities and Society 114:105776. https://doi.org/10.1016/j.scs.2024.105776

Zhang W, Zeng M (2024) Is artificial intelligence a curse or a blessing for enterprise energy intensity? Evidence from China. Energy Economics 134:107561. https://doi.org/10.1016/j.eneco.2024.107561

Zhao Y, Yin D, Wang L, Yu Y (2024) The rise of artificial intelligence, the fall of human wellbeing? International Journal of Social Welfare 33:75–105. https://doi.org/10.1111/ijsw.12586

Zürner S, Deutschländer LP, Schieck M, Franczyk B (2023) Sustainable Development of AI applications in Agriculture: A Review. Procedia Computer Science 225:3546–3553. https://doi.org/10.1016/j.procs.2023.10.350